\documentclass[prd,aps,twocolumn,nofootinbib,preprintnumbers,superscriptaddress,preprintnumbers,balancelastpage,longbibliography]{revtex4-2}
\usepackage[english]{babel}

\usepackage{amsmath,mathtools,physics,xfrac}
\usepackage{amssymb}
\usepackage{graphicx}
\usepackage{afterpage}
\usepackage{float}
\usepackage{subfigure}
\usepackage{rotating}
\usepackage{multirow}
\usepackage{tabularx}
\usepackage{booktabs}
\usepackage{fancyhdr}
\usepackage[utf8]{inputenc}
\usepackage{theorem}
\usepackage{moreverb}
\usepackage{euscript}
\usepackage{psfrag}
\usepackage{slashed}
\usepackage{mathtools}
\usepackage{makecell}
\usepackage{adjustbox}
\usepackage{dcolumn}
\usepackage{bm}
\usepackage[dvipsnames]{xcolor}
\usepackage{graphics}
\usepackage{hyperref}
\usepackage{chngcntr}
\usepackage{aas_macros,orcidlink}
\usepackage{color,xcolor}
\hypersetup{
     colorlinks   = true,
     citecolor    = Aquamarine,
     urlcolor     = Aquamarine,
     linkcolor    = Aquamarine
}

\definecolor{darkblue}{rgb}{0.0,0.0,0.75}
\definecolor{darkred}{rgb}{0.6,0.0,0}
\definecolor{darkgreen}{rgb}{0.0,0.6,0.}

\counterwithout{equation}{section}

\newcommand{\eV}{\,\mathrm{eV}}

\newcommand{\Mpc}{\,\mathrm{Mpc}}

\newcommand{\neff}{N_\mathrm{eff}}

\newcommand{\Om}{\Omega_\mathrm{m}}

\begin{document}
\title{\boldmath Cosmological constraints on mass-varying dark matter}

\author{Amlan Chakraborty\,\orcidlink{0000-0001-9716-7875}}
\email{amlan.chakraborty@iiap.res.in}
\affiliation{Indian Institute of Astrophysics, Bengaluru, Karnataka 560034, India}

\author{Anirban Das\,\orcidlink{0000-0002-7880-9454}}
\email{anirbandas.21@protonmail.com}
\affiliation{Center for Theoretical Physics, Department of Physics and Astronomy, Seoul National University, Seoul 08826, South Korea}
\affiliation{Theory Division, Saha Institute of Nuclear Physics, Kolkata 700064, India}
\affiliation{Homi Bhabha National Institute, Training School Complex, Anushaktinagar, Mumbai
400094, India}

\author{Subinoy Das}
\email{subinoy@iiap.res.in}
\affiliation{Indian Institute of Astrophysics, Bengaluru, Karnataka 560034, India}

\author{Shiv K. Sethi}
\email{sethi@rri.res.in}
\affiliation{Raman Research Institute, C.~V. Raman Avenue, Sadashivanagar, Bengaluru 560080, India}


\begin{abstract}
 
As one of the fundamental unknowns of our Universe, the mass of dark matter remains to be a topic of great interest. We consider the possibility of a time-variation of the dark matter mass.
We study the cosmological constraints on a model where the dark matter mass transitions from zero to a finite value in the early Universe. In this model, the matter power spectrum exhibits power suppression below a certain scale that depends on the epoch of transition, and the angular power spectrum of the cosmic microwave background shows a distinctive phase shift and power suppression at small scales. We use the latest cosmic microwave background data and the $S_8$ priors from weak lensing data to place a lower limit on the transition redshift. 
We also find that the data from the ACT show a mild preference for the mass-varying dark matter model over $\Lambda$CDM.

\end{abstract}
 
\maketitle

\section{Introduction}
Planck satellite observations of the Cosmic Microwave Background (CMB) tell us that dark matter (DM) forms the dominant share of all matter in the Universe\,\cite{refId0}. 
The most successful theory to explain the observations at different scales, ranging from galaxies and galaxy clusters to the horizon, is the $\Lambda$CDM model, which includes a cold, collisionless form of dark matter (CDM). All of these observations are, however, based on the gravitational effects of DM on the visible Universe, and hence, its exact particle nature remains elusive to date.

Numerous laboratory and space-based experiments have been conducted over the past few decades to shed light on the elusive particle nature of DM. However, despite these efforts, no conclusive evidence of non-gravitational interactions with standard model particles has been found yet. In particular, the non-detection of weakly interacting massive particles (WIMPs), arguably considered one of the most promising candidates for DM over the last two decades, has raised concerns \cite{Salucci2019}. This has resulted in a wider search program in the DM model space. It also underscores the importance of a more careful study of the gravitational signatures of DM in cosmology.
Additionally, problems related to the structure formation of the universe at a small scale, collectively known as the diversity problem, have further motivated the exploration of beyond-cold dark matter (CDM) models. While some of these issues have been addressed through baryonic feedback, it is still worthwhile to investigate alternative scenarios beyond CDM for cosmological purposes.

Warm dark matter (WDM) and mixed dark matter, which is a combination of hot dark matter (HDM) and WDM or CDM, have been proposed as alternatives to CDM to solve small-scale problems due to their free-streaming properties, which result in less structure formation at small scales\,\cite{PhysRevD.88.043502,Holm_2022, Blinov_2020, Alexander_2023}. It reduces the number of satellite galaxies in halos and suppresses the halo mass function. However, in addition to these small-scale problems, the standard $\Lambda$CDM model has been challenged by cosmological tensions such as the Hubble tension and the $S_8$ tension\,\cite{Abdalla:2022yfr}.  

The clumpiness of the matter distribution in our Universe is parameterized by $S_8=\sigma_8(\Om/0.3)^{0.5}$, where $\sigma_8$ is the root mean square of matter fluctuations on an $8h^{-1}\Mpc$ scale, and $\Om$ is the total matter abundance. In $\Lambda$CDM, Planck observation yields $S_8= 0.832 \pm 0.013$ \cite{refId0}. The observation of weak lensing of galaxies from the CFHTLenS collaboration initially pointed to a higher $S_8 =0.799 \pm 0.015$ at the $2\sigma$ level \cite{10.1093/mnras/stt601,10.1093/mnras/stv1154} in the framework of $\Lambda$CDM. However, a re-analysis with the combination of DES data \cite{PhysRevD.98.043526} and KiDS/Viking \cite{refId01,refId02} established the tension at the $3\sigma$ level with the value $S_8=0.755^{+0.019}_{-0.021}$ \cite{refId02}. The tension arises due to a lower matter clustering at scales $k\approx 0.1-1 h\Mpc^{-1}$, and a reduction in the amplitude of matter fluctuations on those scales can effectively resolve it. 

Despite numerous attempts, alternative CDM scenarios such as WDM or mixed dark matter models have failed to resolve the $S_8$ tension. Adding massive thermal neutrinos, one of the proposed solutions does not alleviate the tension. This is because an increase in $\Delta N_{\rm eff}$ can raise $\sigma_8$ and lower $\Om$, which exacerbates the tension \cite{refId0,PhysRevD.97.123504}. Various approaches have been explored, such as decaying dark matter \cite{Enqvist_2015,Poulin_2016,PhysRevD.99.121302,10.1093/mnras/staa1991,PhysRevD.103.043014,Pandey_2020,PhysRevD.105.063525,PhysRevD.104.123533} and non-thermal dark matter models \cite{PhysRevD.97.123504,PhysRevD.101.123505,Murgia_2016,Archidiacono_2019,Becker_2021,PhysRevD.94.123511}, to address this issue.

In this paper, we investigate the cosmological signatures of a novel scenario of mass-varying DM (MVDM) that is a natural generalization of the extra radiation $\neff$ and the WDM models. The realisation of an MVDM scenario has been previously suggested in many particle physics models \cite{Agarwal:2014qca, Das:2006ht,Chanda:2017coy, Ganesan:2024bsf}. However, in this work, we refrain from assuming any particular model and assume a phenomenological relation proposed in \cite{Das:2023enn}.  
 
Previously, cosmological effects of late formation of massive DM were studied in Refs.\,\cite{Sarkar:2014bca,Das:2018ons,Das:2023enn,PhysRevD.103.043517}. Similar phenomenon can also occur in models with late primordial black hole formation\,\cite{Bhattacharya:2021wnk, Chakraborty_2022}. Particularly in Ref.\,\cite{Das:2023enn}, some of the authors showed in a model-independent approach that the Lyman-$\alpha$ data from small scale places strong constraint on the transition redshift $z_t$. Here, we follow a similar phenomenological approach to model the time-variation of DM mass from zero to a finite value.
 
We perform a comprehensive Bayesian analysis of this model using the latest cosmic microwave background (CMB) data from Planck and ACT. Our study reveals a lower bound on the transition redshift $z_t$ ($\log_{10} z_t >4.55$) and a mild preference for a DM mass of $41.7$ eV when the Planck \cite{refId0} and KIDS1000+BOSS+2dfLenS weak lensing data \cite{kidsref} are taken into account. The results indicate that the weak lensing data favors this model. Moreover, the inclusion of weak lensing data significantly reduces the $S_8$ tension as compared to the case when only Planck data is considered.

The plan of the paper is as follows. In section \ref{bkg}, we describe the model and the framework of MVDM and discuss its cosmological signatures. In section \ref{method}, we present the method of our Markov Chain Monte Carlo (MCMC) analysis with different datasets and discuss the results in section \ref{results}. Finally, we conclude in section \ref{conclusion}.

\section{Background Density Evolution and Perturbations} \label{bkg}

In this study, we consider MVDM to be a fermionic particle with a temperature $T$. 
We do not commit to a particular particle physics model for the mass variation of DM. Instead, we adopt the phenomenological relation proposed in Ref.\,\cite{Das:2023enn}. The essential concept is that DM was massless and relativistic in the early time, and at a transition redshift of $z_t$, it underwent an instantaneous transition to become non-relativistic. The form of $m(z)$ considered in this analysis is as follows
\begin{equation}\label{eq:dm_mass}
    m(z)= \frac{M}{2} \left[1 - \tanh{\left( \frac{z - z_t}{\Delta  z}\right)} \right]\,.
\end{equation}
\begin{figure}
    \centering
    \includegraphics[scale=0.4]{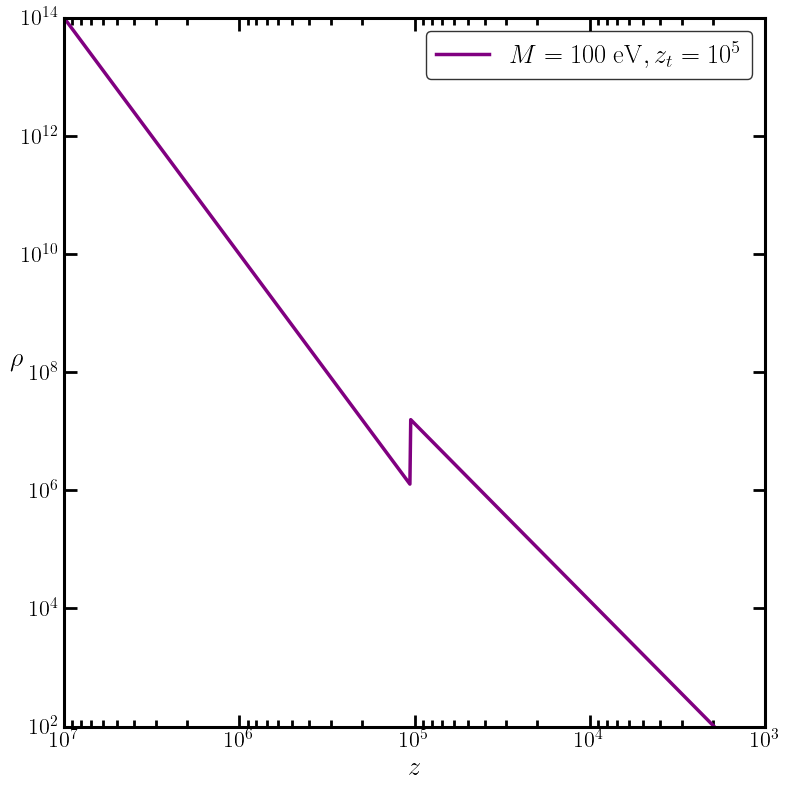}
    \caption{Evolution of background energy density with redshifts is plotted for mass $M=100 \eV$  for a transitional redshift $z_t= 10^5$.}
    \label{fig:bkg_Evolution}
\end{figure}
Here, $M$ denotes the final mass of MVDM, whereas the redshift duration of the transition is represented by $\Delta z$. We note that, due to our assumption of the instantaneous nature of the transition, the value of $\Delta z$ is considerably smaller than that of $z_t$ ($\Delta z\ll z_t$). Consequently, the DM instantaneously loses its relativistic nature, necessitating the ratio of the final mass ($M$) to the transition temperature ($T_t$) to be significantly greater than unity, i.e., $M/T_t \gg 1$.

\begin{figure*}[t]
    \centering
    \includegraphics[scale=0.45]{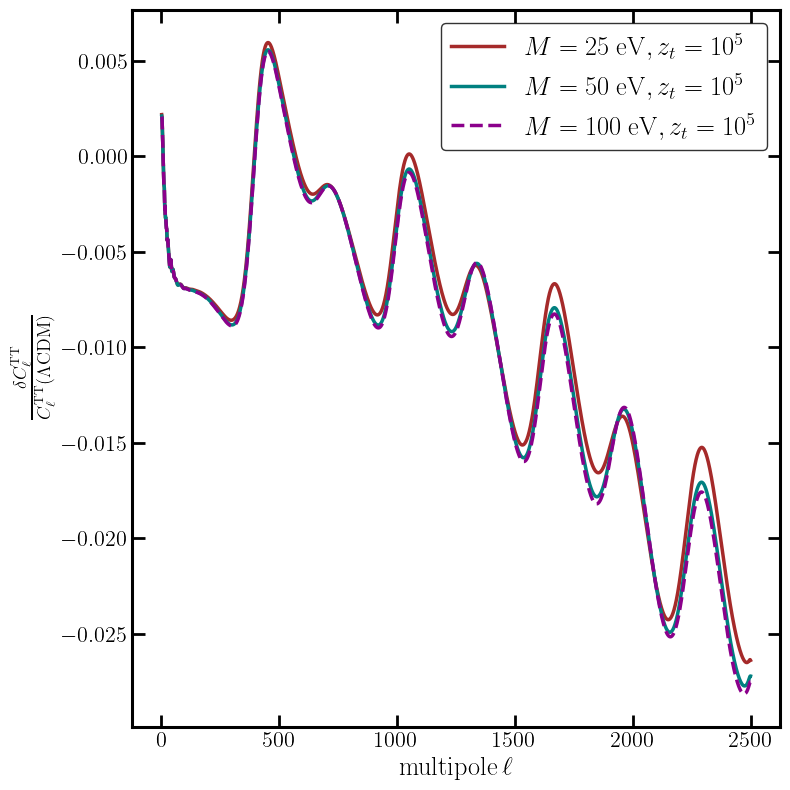}~
    \includegraphics[scale=0.45]{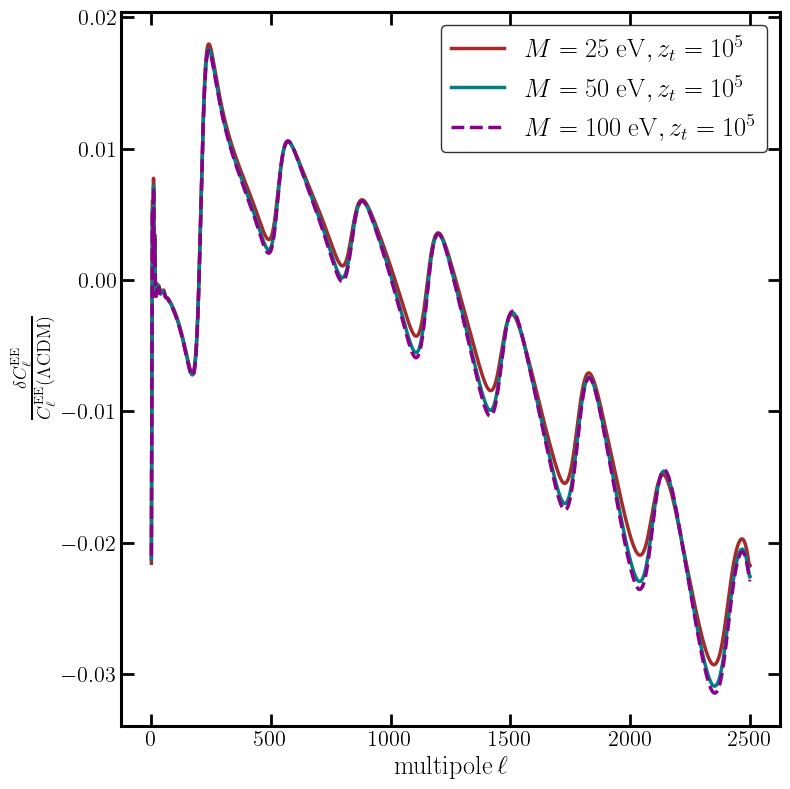}
    \caption{The relative change in the CMB TT and EE power spectra with respect to $\Lambda$CDM is plotted for mass $M= 25,\; 50\; \textrm{and}\; 100 \eV$  for a transitional redshift $z_t= 10^5$. }
    \label{fig:cmb_power}
\end{figure*}

\begin{figure}[t]
    \centering
    \includegraphics[scale=0.42]{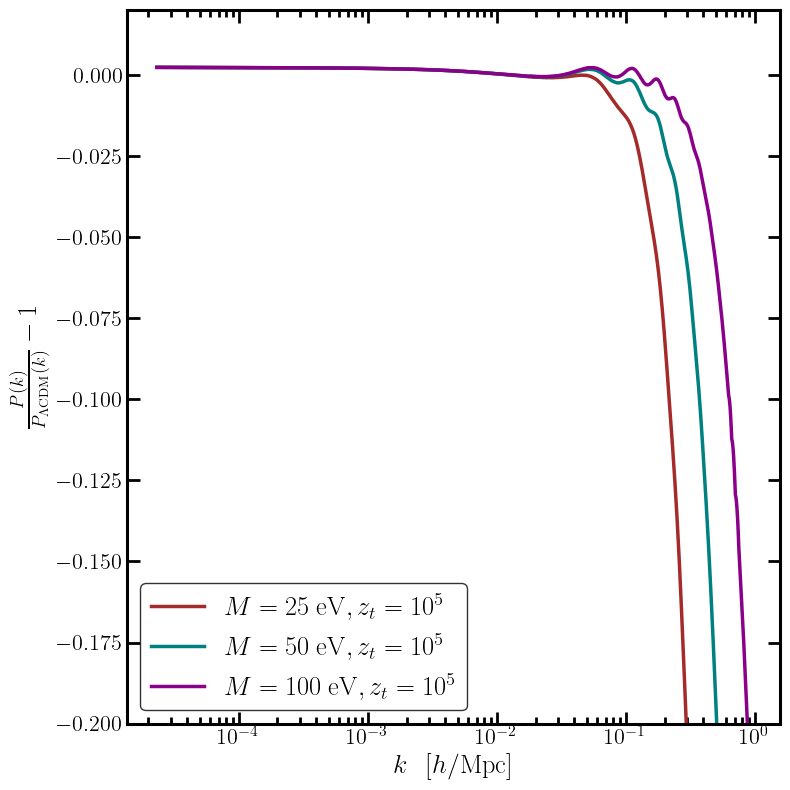}
    \caption{Relative change in matter power spectra with respect to $\Lambda$CDM is plotted for mass $M=25,\; 50\; \textrm{and}\; 100 \eV$  for a transitional redshift $z_t= 10^5$. }
    \label{fig:mpk_100}
\end{figure}

We take the phase space distribution of MVDM to be Fermi-Dirac, which enables us to calculate the energy density and other higher moments of the distribution. Due to the rapid transition, the energy density exhibits a jump during its evolution at $z_t$, which can be expressed as the quantity $ \rho_{\rm MVDM}^{\rm NR}(z\to z^{+}_t/\rho_{\rm MVDM}^{\rm R}(z\to z^{-}_t)=M/T_t$. This phenomenon is accurately depicted in Figure \ref{fig:bkg_Evolution} for a particular value of $z_t$. For the remainder of this paper, we have set the MVDM temperature to $T=T_{\gamma}/10$ in the relativistic phase as a representative value. The mass of a light, Fermionic dark matter particle can be constrained
by Tremaine-Gunn bound\,\cite{peebles:1993}. This bound can be cast in terms of the value of the initial (fine-grained) phase space distribution function (Ref.\,\cite{peebles:1993}, Eq.(18.64)). The constraint depends on the ratio $M/T$, which means the bounds need to be scaled by the corresponding factor for 
our choice of temperature, $T$. We note that the parameter space studied in the paper is consistent with this bound.

Because of its relativistic nature before $z_t$, MVDM will contribute to the total radiation energy density, thereby affecting the evolution of the universe and leaving an imprint on observables such as the CMB and the matter power spectra. This contribution can be quantified as an additional relativistic degree of freedom, $\Delta\neff$, given by \cite{Das:2023enn},

\begin{equation} \label{eq:n_eff}
    \Delta\neff \approx \frac{\rho_{\rm MVDM} (z=0)}{\rho_{\nu}^{\rm th} (z=0)} \frac{T_t}{M} \frac{1}{1+z_t}
\end{equation}

This model shows observable effects at the perturbation level as well. The transition of DM from relativistic to non-relativistic phase introduces a cut-off scale, denoted by $k_t$, in our model, which corresponds to the fluctuations that entered the horizon at the transition redshift $z_t$. Prior to the transition redshift, DM was relativistic and free streaming, prohibiting the growth of structure for the modes that entered before $z_t$. This results in a suppression of power for $k>k_t$ in the matter power spectra shown in Figure \ref{fig:mpk_100}. For lower transition redshift, the suppression moves to larger scales, implying the structure is washed away at lower $k$ modes. Furthermore, the free streaming length is directly proportional to the thermal velocity of DM particles, i.e., $v_{th}= \frac{\langle p \rangle}{m}$. Therefore, the cut-off scale $k_t$, which is determined from the free-streaming length during the transition, is also dependent on the final mass of the DM\,\cite{Hu:1997mj,Lesgourgues:2006nd,Viel:2013fqw,TopicalConvenersKNAbazajianJECarlstromATLee:2013bxd}. Consequently, we obtain smaller $k_t$ values for DM with lighter mass. This suppression of power is also reflected in the CMB temperature and polarization anisotropy power spectrum at higher $\ell$, as shown in Figure \ref{fig:cmb_power}.

The presence of free-streaming dark matter (DM) particles during their relativistic phase causes a small but non-zero $\Delta \neff$, which exerts an additional drag on the metric perturbation via gravity during the radiation-dominated era. This results in an extra phase shift in the acoustic oscillation of the photon-baryon plasma that can be detected in both the cosmic microwave background (CMB) and the matter power spectrum as small wiggles, which can be seen in both Figure \ref{fig:cmb_power} \& \ref{fig:mpk_100}.

\begin{figure*}
\includegraphics[scale=0.6]{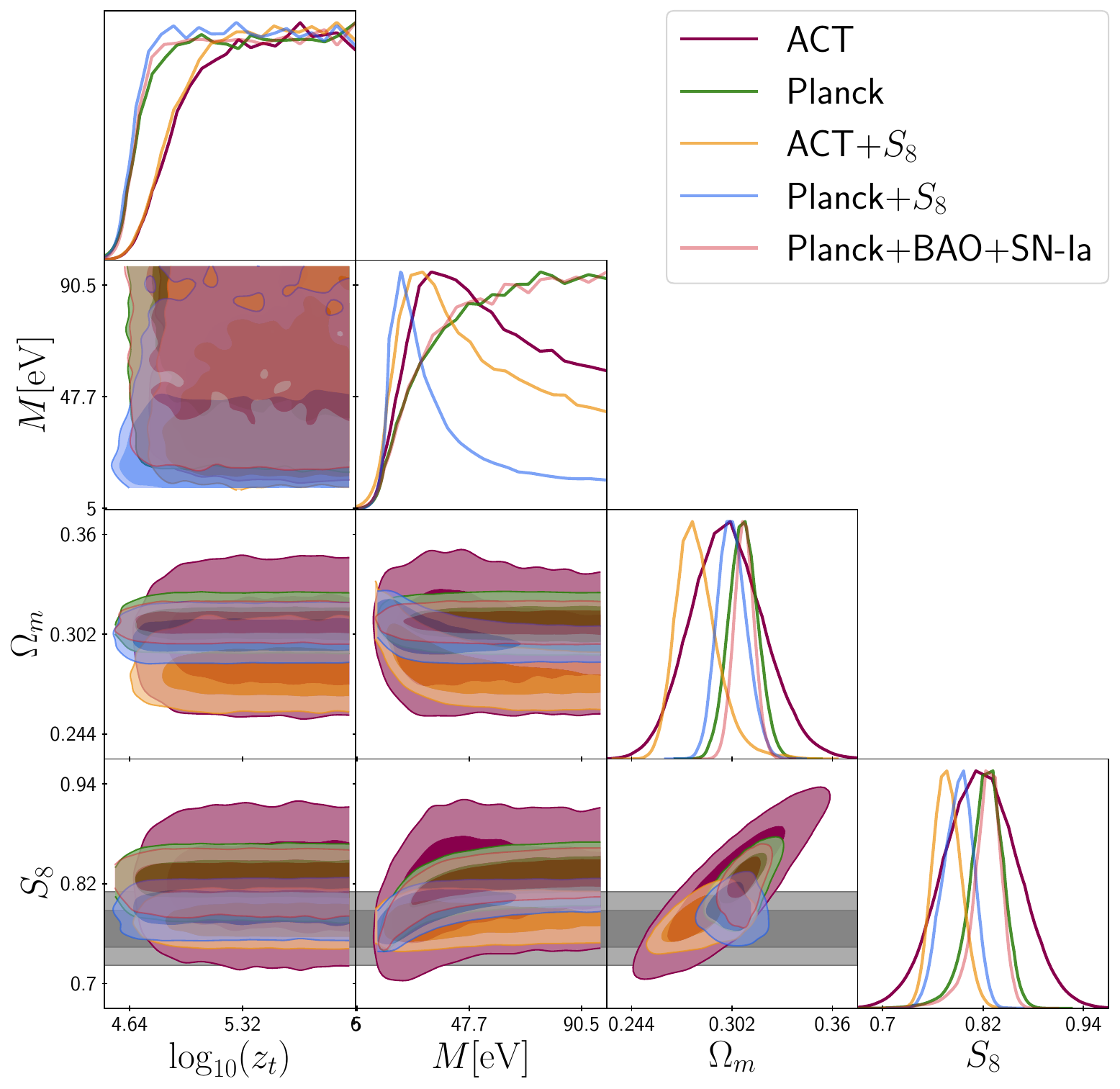}
\caption{Reconstructed 2D and 1D marginalized posterior distributions of ${\log_{10} z_t, M(\eV), \Om, S_8}$ with $68\%$ and $95\%$ confidence level. The weak lensing measurement of $S_8$ is shown as grey bands (68\% and 95\%). Here we considered Pantheon+ dataset for "SN-Ia".}\label{fig:mcmc_results}
\end{figure*}

\begin{figure*}
\includegraphics[scale=0.6]{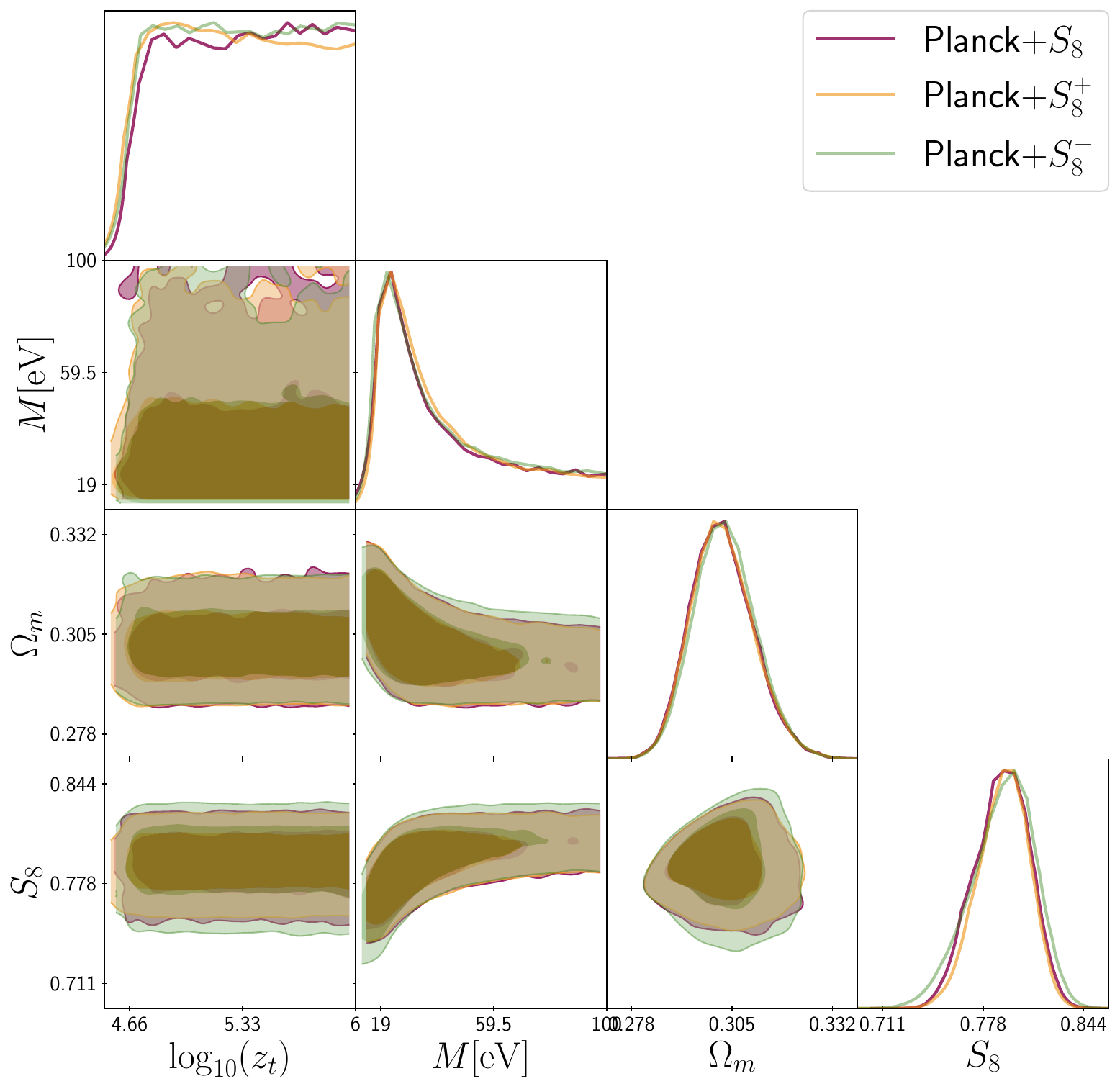}
\caption{The 2D and 1D marginalized posterior distributions of ${\log_{10} z_t, M, \Om, S_8}$ with $68\%$ and $95\%$ confidence level for three different $S_8$ priors as mentioned in Sec.\,\ref{method}.}\label{fig:mcmc_diff_S8_results}
\end{figure*}

\begin{table*}
\caption{\label{tab:table1} The mean (best-fit) $\pm 1\sigma$ error of the cosmological parameters in the $\Lambda$CDM and MVDM model obtained from the analysis of Planck \cite{refId0} and Planck+$S_8$\cite{kidsref} data. Lower limits are obtained at half of the peak posterior value of the respective parameter. }
\begin{ruledtabular}
\begin{tabular}{ccccc}
 Model&\multicolumn{2}{c}{$\Lambda$CDM}&\multicolumn{2}{c}{MVDM}\\[1ex] \hline\rule{0pt}{1.2\normalbaselineskip}
 Parameter&Planck&Planck+$S_8$&Planck
&Planck+$S_8$\\[1ex] \hline
\rule{0pt}{1.2\normalbaselineskip}
$100\omega_b$&$2.24(2.2334)^{+0.0154}_{-0.0155}$&$2.25(2.2528)^{+0.0145}_{-0.0146}$&$2.24(2.239)^{+0.0153}_{-0.0157}$&$2.25(2.245)^{+0.0152}_{-0.0156}$ \\[1ex]
 $\omega_{\rm dm}\footnote{for $\Lambda$CDM model, the 'dm' subscript means the Cold Dark Matter (CDM), and for the MVDM model, it means the mass varying dark matter considered in this paper.}$&$0.12(0.1199)^{+0.00141}_{-0.0014}$&$0.118(0.1178)^{+0.00114}_{-0.00118}$&$0.12(0.1208)^{+0.00141}_{0.00143}$&$0.119(0.12)^{+0.00134}_{-0.00158}$\\[1ex]
 $100\times\theta_s$&$1.04(1.04184)^{+0.000304}_{-0.0003}$&$1.04(1.04215)^{+0.000301}_{-0.000294}$
 &$1.04(1.042)^{+0.000322}_{-0.000353}$&$1.04(1.0419)^{+0.000317}_{-0.000362}$ \\[1ex]
 $n_s$&$0.965(0.96754)\pm 0.00453$&$0.969(0.972)^{+0.00421}_{-0.00422}$&$0.968(0.9655)^{+0.00466}_{-0.00503}$&$0.97(0.965)^{+0.00493}_{-0.00511}$\\[1ex]
 $\ln{10^{10}A_s}$&$3.04(3.0391)^{+0.0156}_{-0.0162}$&$3.04(3.021)^{+0.0159}_{-0.0158}$&$3.05(3.0442)^{+0.0162}_{-0.0173}$&$3.04(3.042)^{+0.0158}_{-0.0171}$\\[1ex]
 $\tau_{reio}$&$0.0541(0.0517)^{+0.00763}_{-0.00786}$&$0.0521(0.0507)^{+0.00799}_{-0.00785}$&$0.0546(0.0547)^{+0.00786}_{-0.00847}$&$0.0538(0.05198)^{+0.00758}_{-0.00826}$ \\[1ex]
 $\log_{10}z_t$&---&---&$>4.72$&$>4.67$ \\[1ex]
 $M$[eV]&---&---&$>23$&$41.7(23.87)^{+7.81}_{-27.5}$ \\[1ex] \hline\rule{0pt}{1.2\normalbaselineskip}
 $S_8$&$0.831(0.827)^{+0.0164}_{-0.0165}$&$0.805(0.8114)^{+0.0126}_{-0.0129}$&$0.822(0.821)^{+0.0236}_{-0.0179}$ &$0.79(0.776)^{+0.0209}_{-0.0167}$ \\[1ex]
 $\Om$&$0.315(0.3138)^{+0.00852}_{-0.00867}$&$0.303(0.3064)^{+0.00672}_{-0.00712}$&$0.309(0.301)^{+0.00823}_{-0.00884}$&$0.302(0.311)^{+0.00759}_{-0.0095}$ \\[1ex]
 $H_0$[Km/s/Mpc]&$67.3(67.35)^{+0.602}_{-0.643}$&$68.2(67.88)^{+0.528}_{-0.527}$&$67.9(67.52)^{+0.638}_{-0.636}$&$68.4(67.96)^{+0.695}_{-0.606}$ \\[1ex] \hline
 \rule{0pt}{1.2\normalbaselineskip}
 $\chi^2_{\rm min}$ &$2771.78$&$2778.14$&$2770.55$&$2774.33$ \\
\end{tabular}
\end{ruledtabular}
\end{table*}

\begin{table*}
\caption{\label{tab:table2} The mean (best-fit) $\pm 1\sigma$ error of the cosmological parameters in the $\Lambda$CDM and MVDM model obtained from the analysis of ACT \cite{Aiola_2020, Louis_2017} and ACT+$S_8$\cite{kidsref} data. Lower limits are obtained at half of the peak posterior value of the respective parameter. }
\begin{ruledtabular}
\begin{tabular}{ccccc}
 Model&\multicolumn{2}{c}{$\Lambda$CDM}&\multicolumn{2}{c}{MVDM}\\[1ex] \hline\rule{0pt}{1.2\normalbaselineskip}
 Parameter&ACT&ACT+$S_8$&ACT
&ACT+$S_8$\\[1ex] \hline
\rule{0pt}{1.2\normalbaselineskip}
$100\omega_b$&$2.15(2.1471)^{+0.0308}_{-0.0318}$&$2.16(2.1561)^{+0.0307}_{-0.0314}$&$2.16(2.166)\pm 0.0317$&$2.17(2.1698)^{+0.0304}_{-0.0323}$ \\[1ex]
 $\omega_{\rm dm}\footnote{for $\Lambda$CDM model, the 'dm' subscript means the Cold Dark Matter (CDM), and for the MVDM model, it means the mass varying dark matter considered in this paper.}$&$0.118(0.1171)^{+0.00373}_{-0.00384}$&$0.114(0.11401)^{+0.00162}_{-0.00177}$&$0.118(0.11839)^{+0.00379}_{-0.00385}$&$0.115(0.1168)^{+0.00179}_{-0.00266}$\\[1ex]
 $100\times\theta_s$&$1.04(1.0433)^{+0.000728}_{-0.000708}$&$1.04(1.04385)^{+0.000666}_{-0.000686}$
 &$1.04(1.0436)^{+0.000723}_{-0.000727}$&$1.04(1.0436)^{+0.000691}_{-0.000698}$ \\[1ex]
 $n_s$&$1.01(1.0092)^{+0.0156}_{-0.0159}$&$1.02(1.0177)^{+0.0142}_{-0.0144}$&$1.01(1.00945)^{+0.0154}_{-0.0164}$&$1.02(1.0131)^{+0.0147}_{-0.0149}$\\[1ex]
 $\ln{10^{10}A_s}$&$3.04(3.04165)^{+0.0224}_{-0.0238}$&$3.03(3.0332)^{+0.0209}_{-0.0223}$&$3.04(3.0421)^{+0.0225}_{-0.0239}$&$3.03(3.029)^{+0.0209}_{-0.0232}$\\[1ex]
 $\tau_{reio}$&$0.0604(0.06058)^{+0.00927}_{-0.01}$&$0.0601(0.06145)^{+0.00897}_{-0.0102}$&$0.0606(0.05936)^{+0.00922}_{-0.0103}$&$0.0604(0.05645)^{+0.00914}_{-0.0102}$ \\[1ex]
 $\log_{10}z_t$&---&---&$>5.1$&$>5.2$ \\[1ex]
 $M$[eV]&---&---&$>27$&$51.2(25.31)^{+16}_{-33.5}$ \\[1ex] \hline\rule{0pt}{1.2\normalbaselineskip}
 $S_8$&$0.826(0.8169)^{+0.0419}_{-0.044}$&$0.777(0.78)^{+0.016}_{-0.0183}$&$0.817(0.8159)^{+0.0452}_{-0.046}$ &$0.777(0.7711)^{+0.0159}_{-0.0183}$ \\[1ex]
 $\Om$&$0.303(0.2986)^{+0.0202}_{-0.023}$&$0.279(0.2822)^{+0.00842}_{-0.00955}$&$0.3(0.2998)^{+0.0203}_{-0.0226}$&$0.283(0.2913)^{+0.00916}_{-0.0143}$ \\[1ex]
 $H_0$[Km/s/Mpc]&$67.9(68.11)^{+1.51}_{-1.55}$&$69.6(69.384)^{+0.736}_{-0.699}$&$68.4(68.339)^{+1.51}_{-1.58}$&$69.6(68.95)^{+1.06}_{-0.768}$ \\[1ex] \hline
 \rule{0pt}{1.2\normalbaselineskip}
 $\chi^2_{\rm min}$ &$280.0495$&$281.8939$&$275.7832$&$276.7859$ \\
\end{tabular}
\end{ruledtabular}
\end{table*}

\begin{table*}
\caption{\label{tab:table3} The mean (best-fit) $\pm 1\sigma$ error of the cosmological parameters in the $\Lambda$CDM and MVDM model obtained from the analysis of combined Planck + BAO + Pantheon+ dataset. Lower limits are obtained at half of the peak posterior value of the respective parameter.}
\begin{ruledtabular}
\begin{tabular}{ccc}
 Dataset&\multicolumn{2}{c}{Planck + BAO + Pantheon+}\\[1ex] \hline\rule{0pt}{1.2\normalbaselineskip}
 Parameter&$\Lambda$CDM&MVDM \\[1ex] \hline
\rule{0pt}{1.2\normalbaselineskip}
$100\omega_b$&$2.24(2.2403)^{+0.0136}_{-0.0138}$&$2.24(2.249)^{+0.0137}_{-0.0140}$ \\[1ex]
 $\omega_{\rm dm}\footnote{for $\Lambda$CDM model, the 'dm' subscript means the Cold Dark Matter (CDM), and for the MVDM model, it means the mass varying dark matter considered in this paper.}$&$0.12(0.1201)^{+0.000981}_{-0.000997}$&$0.121(0.1186)^{+0.000982}_{-0.000964}$\\[1ex]
 $100\times\theta_s$&$1.04(1.0419)^{+0.000284}_{-0.000289}$&$1.04(1.0420)^{+0.000311}_{-0.000342}$ \\[1ex]
 $n_s$&$0.966(0.9645)^{+0.00376}_{-0.00387}$&$0.967(0.9618)^{+0.00387}_{-0.00427}$\\[1ex]
 $\ln{10^{10}A_s}$&$3.04(3.038)^{+0.0161}_{-0.0168}$&$3.05(3.045)^{+0.0157}_{-0.0166}$\\[1ex]
 $\tau_{reio}$&$0.0546(0.051609)^{+0.00756}_{-0.00809}$&$0.0547(0.05589)^{+0.00755}_{-0.00798}$ \\[1ex]
 $\log_{10}z_t$&---&$>4.75$ \\[1ex]
 $M$[eV]&---&$>26$ \\[1ex] \hline\rule{0pt}{1.2\normalbaselineskip}
 $S_8$&$0.826(0.828)^{+0.0127}_{-0.0128}$&$0.824(0.819)^{+0.0203}_{-0.0132}$ \\[1ex]
 $\Om$&$0.312(0.315)^{+0.00591}_{-0.00605}$&$0.309(0.301)^{+0.00579}_{-0.00588}$ \\[1ex]
 $H_0$[Km/s/Mpc]&$67.5(67.408)^{+0.442}_{-0.441}$&$67.9(68.45)^{+0.427}_{-0.443}$ \\[1ex] \hline
 \rule{0pt}{1.2\normalbaselineskip}
 $\chi^2_{\rm min}$ &$4189.92$&$4189.66$ \\
\end{tabular}
\end{ruledtabular}
\end{table*}

\begin{table*}[ht]
\caption{\label{tab:table_aic} Comparison of $\Delta \chi^2_{\rm min}$ and $\Delta$AIC per experiment for MVDM and $\Lambda$CDM models.
}
\begin{ruledtabular}
\begin{tabular}{lccc}
\textbf{Dataset} & \multicolumn{2}{c}{MVDM}  \\[1ex]
\hline\rule{0pt}{1.2\normalbaselineskip}
 & $\Delta \chi^2_{\rm min}$ & $\Delta$AIC  \\[1ex]
\hline\rule{0pt}{1.2\normalbaselineskip}

Planck & -1.23 & +2.77 \\[1ex]
Planck+$S_8$ & -3.81 & +0.19  \\[1ex]
ACT & -4.2663 & -0.2663  \\[1ex]
ACT+$S_8$ & -5.108 & -1.108  \\[1ex]
Planck+BAO+SN-Ia & -0.26 & +3.74  \\[1ex]

\end{tabular}
\end{ruledtabular}
\end{table*}

\begin{figure*}[!t]
\includegraphics[scale=0.55]{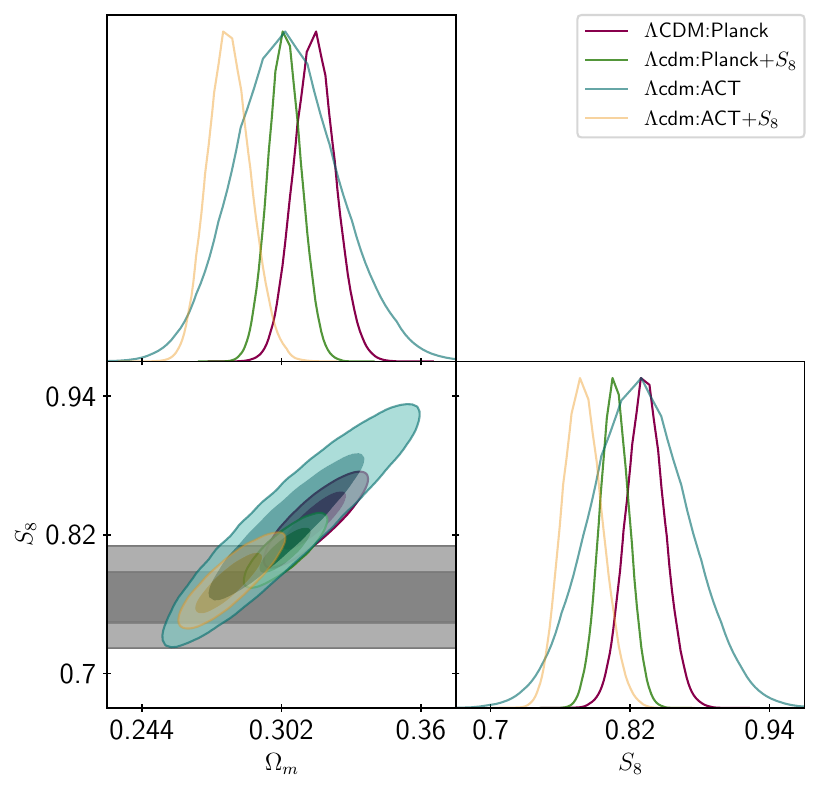}~\qquad
\includegraphics[scale=0.55]{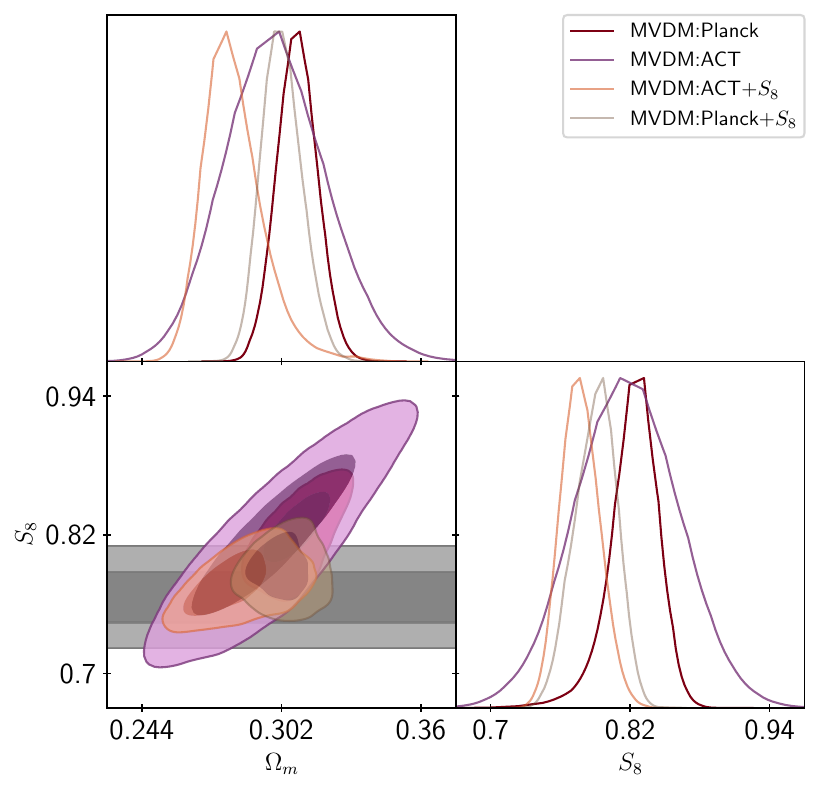}
\caption{Reconstructed 2D marginalized posterior distributions of $\Om$ and $S_8$ with $68\%$ and $95\%$ confidence levels.}\label{fig:all s_8_contour_results}
\end{figure*}

\begin{table*}
\caption{\label{tab:table4} The mean (best-fit) $\pm 1\sigma$ error of the cosmological parameters in the  MVDM model obtained from the analysis of Planck+$S_8^+$ \cite{PhysRevD.105.023514}, Planck+$S_8$\cite{kidsref} and Planck+$S_8^-$ \cite{refId0kid} data. Lower limits are obtained at half of the peak posterior value of the respective parameter. }
\begin{ruledtabular}
\begin{tabular}{cccc}
 Model&\multicolumn{3}{c}{MVDM}\\[1ex] \hline\rule{0pt}{1.2\normalbaselineskip}
 Parameter&Planck+$S_8^+$&Planck+$S_8$&Planck+$S_8^-$\\[1ex] \hline
\rule{0pt}{1.2\normalbaselineskip}
$100\omega_b$&$2.25(2.2527)^{+0.0156}_{-0.0153}$&$2.25(2.245)^{+0.0152}_{-0.0156}$&$2.25(2.247)^{+0.0153}_{-0.0154}$ \\[1ex]
 $\omega_{\rm dm}\footnote{for $\Lambda$CDM model, the 'dm' subscript means the Cold Dark Matter (CDM), and for the MVDM model, it means the mass varying dark matter considered in this paper.}$&$0.119(0.1189)^{+0.00132}_{-0.00158}$&$0.119(0.12)^{+0.00134}_{-0.00158}$&$0.119(0.1194)^{+0.00133}_{0.00153}$\\[1ex]
 $100\times\theta_s$&$1.04(1.0422)^{+0.000329}_{-0.000359}$&$1.04(1.0419)^{+0.000317}_{-0.000362}$
 &$1.04(1.0418)^{+0.000323}_{-0.000344}$ \\[1ex]
 $n_s$&$0.97(0.97133)^{+0.00503}_{-0.00513}$&$0.97(0.965)^{+0.00493}_{-0.00511}$&$0.97(0.9722)^{+0.00484}_{-0.00503}$\\[1ex]
 $\ln{10^{10}A_s}$&$3.04(3.0499)^{+0.0159}_{-0.0168}$&$3.04(3.042)^{+0.0158}_{-0.0171}$&$3.04(3.0568)^{+0.0161}_{-0.0166}$\\[1ex]
 $\tau_{reio}$&$0.0535(0.0564)^{+0.00773}_{-0.00811}$&$0.0538(0.05198)^{+0.00758}_{-0.00826}$&$0.0536(0.0595)^{+0.00777}_{-0.008}$ \\[1ex]
 $\log_{10}z_t$&$>4.72$&$>4.67$&$>4.75$ \\[1ex]
 $M$[eV]&$41.6(26.59)^{+7.26}_{-26.9}$&$41.7(23.87)^{+7.81}_{-27.5}$&$41.6(23.5)^{+8.51}_{-28.4}$ \\[1ex] \hline\rule{0pt}{1.2\normalbaselineskip}
 $S_8$&$0.791(0.787)^{+0.0191}_{-0.0155}$&$0.79(0.776)^{+0.0209}_{-0.0167}$&$0.79(0.782)^{+0.0248}_{-0.0178}$ \\[1ex]
 $\Om$&$0.302(0.3018)^{+0.00763}_{-0.00935}$&$0.302(0.311)^{+0.00759}_{-0.0095}$&$0.303(0.305)^{+0.00783}_{-0.00906}$ \\[1ex]
 $H_0$[Km/s/Mpc]&$68.4(68.46)^{+0.685}_{-0.603}$&$68.4(67.96)^{+0.695}_{-0.606}$&$68.4(68.12)^{+0.664}_{-0.614}$ \\[1ex] 
\end{tabular}
\end{ruledtabular}
\end{table*}

\begin{figure*}[ht]
\includegraphics[scale=0.40]{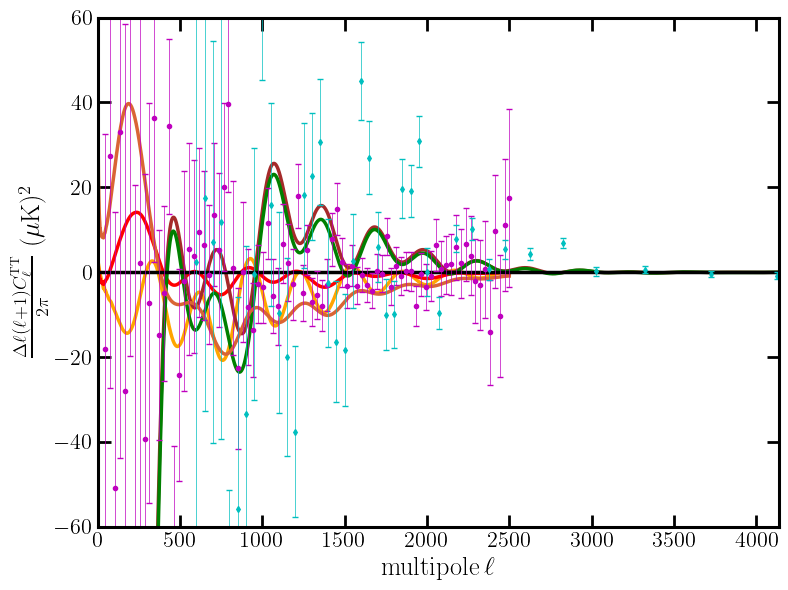}~
\includegraphics[scale=0.41]{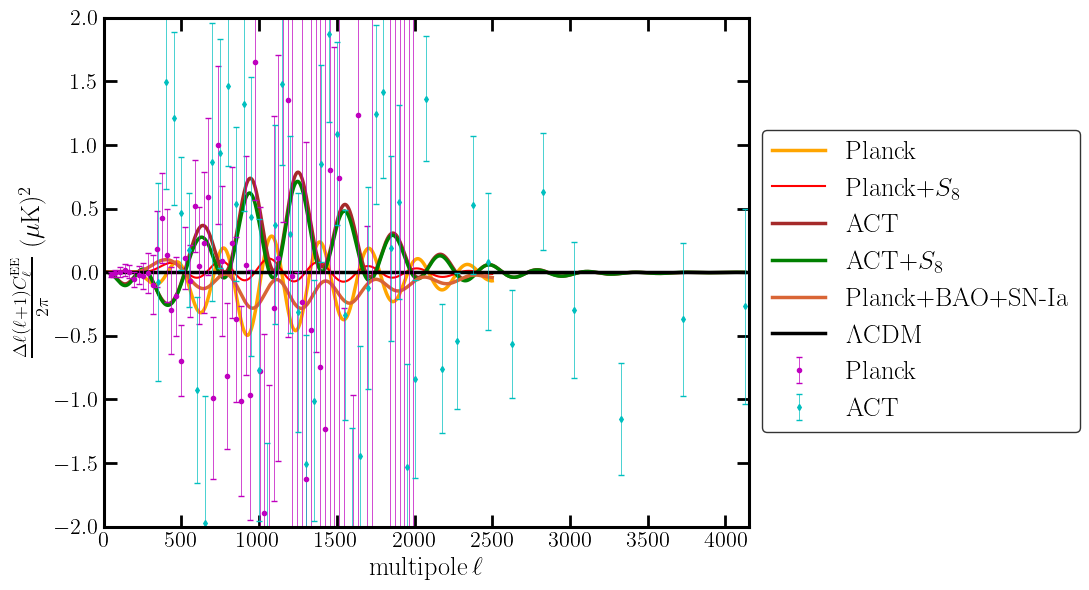}
\includegraphics[scale=0.5]{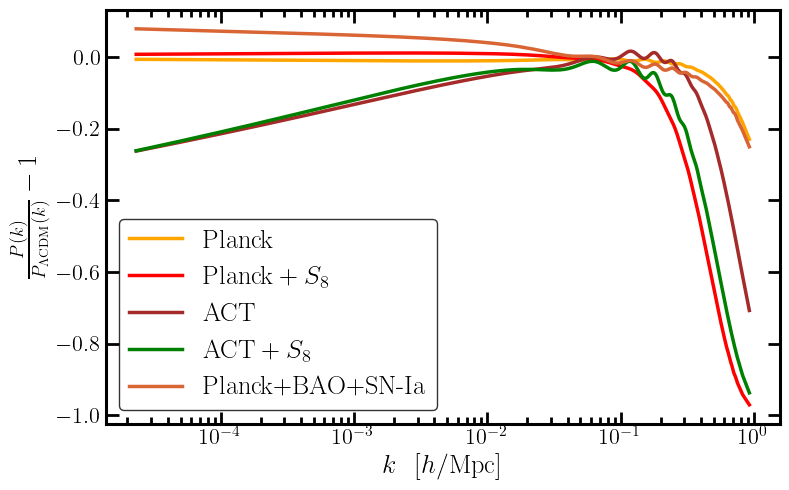}
\caption{\label{fig:residula_plots}Residuals of the CMB TT, EE (top panel) and matter (bottom panel) power spectra with respect to $\Lambda$CDM in the best fit MVDM model for Planck, Planck+$S_8$, ACT, ACT+$S_8$ and Planck+BAO+SN-Ia data set. Here we considered Pantheon+ dataset for "SN-Ia".}
\end{figure*}

\section{Method}\label{method}
We use the Boltzmann hierarchy for the MVDM model implemented in CLASS in Ref.\,\cite{Das:2023enn} to compute the CMB anisotropy power spectra and the matter power spectrum\,\cite{lesgourgues2011cosmic}.

We perform a comprehensive Markov Chain Monte Carlo (MCMC) analysis of the MVDM model using the following combinations of cosmological datasets:
\begin{itemize}
    \item[$\blacktriangleright$] Planck 2018 measurements of the low-$\ell$ CMB TT, EE, and high-$\ell$ TT, TE, EE power spectra \cite{refId0}.  

    \item[$\blacktriangleright$] The BAO measurements from 6dFGS \cite{10.1111/j.1365-2966.2011.19250.x} and SDSS DR7 MGS \cite{10.1093/mnras/stv154} for low redshift $z<0.2$ comprises the likelihood for "Bao-smallz-2014". In addition to this we also use BOSS DR12 at $z=0.38, 0.51$ and $0.61$ \cite{10.1093/mnras/stx721}.  
        
    \item[$\blacktriangleright$] We use the $S_8$-prior from the KIDS1000+BOSS+2dfLenS weak lensing data with $S_8= 0.766^{+0.02}_{-0.014}$\,\cite{kidsref}. In this work, we use the $S_8$ value obtained as a prior for our MCMC analysis. In order to gain an understanding of the effect of prior, we also consider different $S_8$ prior values obtained from different weak lensing measurements. We use a higher $S_8$ prior i.e. $S_8^+=0.772^{+0.018}_{-0.017}$ obtained from shear-shear analysis from DES Y3 survey \cite{PhysRevD.105.023514} as well as a lower prior $S_8^-= 0.759^{+0.024}_{-0.021}$ obtained from KIDS1000-cosmic shear analysis \cite{refId0kid}.
    \item[$\blacktriangleright$] ACT data from ACT collaboration, which includes the DR4 data release \cite{Aiola_2020}. We use \texttt{actpollite dr4} likelihood code which contains TT measurements $(600<\ell<4126)$ and  TE and EE measurements $(350<\ell<4126)$. Since it does not have large-scale data, constraining reionization optical depth $\tau_{\rm reio}$ is not possible. So a Gaussian prior is also added, centered at $\tau_{\rm reio}=0.06\pm 0.01$ while analyzing ACT data alone \cite{Louis_2017}.
    \item[$\blacktriangleright$] Type-Ia supernova (SN-Ia) catalogue Pantheon+ spanning in redshift $0.001<z<2.26$. \cite{Brout_2022}
\end{itemize}
In this study, we adopt the standard $\Lambda$CDM model as our base framework, which comprises of the six parameters: ${\omega_b, \omega_{\rm cdm}, 100\theta_s, \ln{10^{10}A_s},n_s, \tau_{\rm reio}}$. To incorporate our MVDM model, we introduce two additional parameters ${\log_{10} z_t, \mathrm{M}}$ in conjunction with the six aforementioned parameters.

 The MCMCs are executed using the MontePython-v3 \cite{brinckmann2018montepython} code with the Metropolis-Hasting algorithm. We utilize our modified CLASS version to interface with the MCMCs. We obtain all the reported $\chi^2_{min}$ values using the Python package iMinuit \cite{JAMES1975343}. We use a Choleski decomposition to handle various nuisance parameters \cite{Lewis_2000} better and consider chains to be converged when the Gelman-Rubin convergence criterium $R-1\lesssim 0.01$ \cite{10.1214/ss/1177011136}.

\section{Results}\label{results}

For the Planck-only analysis with the MVDM cosmology, shown in Figure \ref{fig:mcmc_results}, we were able to obtain a bound on the mass $M>23\eV$ and $\log_{10} z_t>4.72$ (obtained at half of the peak posterior value of the respective parameter). Our analysis yields $S_8$(MVDM)$=0.822^{+0.0236}_{-0.0179}$, which is to be compared with $S_8(\Lambda$CDM)$= 0.831^{+0.0164}_{-0.0165}$. MVDM shifts $S_8$ downward almost by $\sim 0.5\sigma$, thus effectively reducing the level of $S_8$ tension from $\sim 2.7\sigma$ to $\sim 2.2\sigma$. So, MVDM prefers a slightly lower value of $S_8$, and the minimum $\chi^2$ shows a slight improvement compared to the standard $\Lambda$CDM model i.e. $\Delta \chi^2_{\rm min} = \chi^2_{\rm min} (MVDM) - \chi^2_{\rm min} (\Lambda CDM) = -1.23$.

The inclusion of the $S_8$ prior in our analysis yields significant changes in the results. Specifically, we observe a mild preference for dark matter final mass $M=41.7^{+7.81}_{-27.5}\eV$ and a tighter constraint on $\log_{10} z_t>4.67$ in the MVDM model. We get $S_8$(MVDM)$=0.79^{+0.0209}_{-0.0167}$ in this likelihood combination, which can be compared with our baseline $S_8(\Lambda$CDM)$=0.805^{+0.0126}_{-0.0129}$. Consequently, the $\chi^2_{min}$ in the combined analysis decreases for the MVDM case, with $\Delta \chi^2_{\rm min} = \chi^2_{\rm min} (MVDM) - \chi^2_{\rm min} (\Lambda CDM) = -3.81$. Further, we note that the overall $\chi^2_{\rm min}$ is less impacted by the inclusion of the $S_8$ prior in the MVDM case $(+3.78)$ as compared to that in the $\Lambda$CDM case $(+6.36)$. This finding indicates that the MVDM model is better suited to explain both CMB and weak lensing data compared to $\Lambda$CDM. We also report the individual contribution from each dataset in each run for both of the models, which are reported in Table \ref{tab:table_1} in Appendix \ref{appx:chi2}.

The analysis of Planck data reveals comparatively higher values of $S_8$ , indicating an increased amplitude of matter fluctuations at the $8h^{-1}$Mpc scale. In contrast, the MVDM model exhibits a suppression of fluctuations at smaller length scales, resulting in a reduced amplitude of matter fluctuations. Consequently, the incorporation of $S_8$ priors significantly affects the posterior distributions, demonstrating a mild preference for detecting the mass of dark matter, an aspect that the Planck data alone does not exhibit. Given the substantial impact on the result of MCMC analysis, we performed an additional set of analyses utilizing different $S_8$ priors derived from various weak lensing measurements, as elaborated in Section \ref{method}. The outcomes of this analysis are illustrated in Figure \ref{fig:mcmc_diff_S8_results}. Notably, the posterior distributions indicate that variations in the priors do not lead to significant changes in the results, which is why the minimum $\chi^2$ has not been computed.

To gain a deeper understanding of the results obtained from the MCMC analysis, we present residual plots in Figure \ref{fig:residula_plots} for the CMB TT, EE power spectra, and the matter power spectra with respect to $\Lambda$CDM in the best-fit mass-varying dark matter (MVDM) model obtained when Planck, Planck+$S_8$, and Planck+BAO+Pantheon+ datasets are considered. Notably, the suppression of power at small length scales (i.e., large $k$ modes and large $\ell$'s) is a crucial factor in mitigating the tension. 

As mentioned in Section \ref{bkg} and \cite{Das:2023enn}, this suppression is attributed to the fact that the MVDM stays relativistic until the transition redshift $z_t$, contributing to the relativistic energy density, which can be quantified through additional degrees of freedom in Equation \ref{eq:n_eff}. This effect washes away all perturbations at very small length scales, as evidenced by the matter power spectra residual plot and CMB TT, EE residual plots in Figure \ref{fig:residula_plots}. Consequently, we observe a reduction in the $S_8$ value relative to Planck due to the reduction in fluctuation amplitude on scales $k \approx 0.1-1 h\Mpc^{-1}$ as discussed in the Introduction. The addition of $S_8$ prior yields a mild preference for the mass of MVDM at around $23.87 \eV$ (best-fit value), as shown in Figure \ref{fig:mcmc_results}. These results clearly demonstrate that the KIDS1000+BOSS+2dfLenS weak lensing data mildly prefers the MVDM model.  It should be noted that the error bars in the angular power spectra from Planck are larger at higher $\ell$ owing to the Planck beam, which makes it difficult to put a stringent bound on the Dark Matter mass.  This is because significant deviations from the standard $\Lambda$CDM paradigm occur only at small length scales, as is evident from the matter power spectra residual plot. 

To address this issue, we have also done our MCMC analysis using ACT data, which has better angular resolution than Planck and has measured the CMB power spectra up to $\ell \simeq 4000$. The results are also included in Figure \ref{fig:mcmc_results} and in Table \ref{tab:table2}, where it is clear that the ACT data does a slightly better job in constraining the mass of dark matter and lowering the $S_8$ value than Planck. ACT data provides a lower limit on the mass $M>27\eV$ and $\log_{10} z_t>5.1$ (obtained at half of the peak posterior values of the respective parameters). Our analysis also shows $S_8$(MVDM)$=0.817^{+0.0452}_{-0.046}$, compared to $S_8$($\Lambda$CDM)$=0.826^{+0.0419}_{-0.044}$ where MVDM lowers $S_8$ by $~0.2\sigma$. We also observe an improvement in minimum $\chi^2$ over $\Lambda$CDM, i.e. $\Delta \chi^2_{\rm min} = \chi^2_{\rm min} (MVDM) - \chi^2_{\rm min} (\Lambda CDM) = -4.2663$.

In addition to the $S_8$ value, we find a tighter lower bound on $\log_{10} z_t >5.2$. It also tightens the constraint on the dark matter mass $M=51.2\,(25.31)^{+16}_{-33.5}\eV$, compared to previous ACT analysis, but this constraint is much weaker than that of Planck+$S_8$. We find $S_8$(MVDM)$=0.777^{+0.0159}_{-0.0183}$ in comparison with $S_8$($\Lambda$CDM)$=0.777^{+0.016}_{-0.0183}$, showing no significant change in the $S_8$ values in either of the models except for their best-fit values, as given in Table \ref{tab:table2}. However, the addition of prior reduces the $S_8$ tension significantly compared to all the previous analyses considered in this work. For example, $\chi^2_{min}$ in the combined analysis decreases for the MVDM case, with $\Delta \chi^2_{\rm min} = \chi^2_{\rm min} (MVDM) - \chi^2_{\rm min} (\Lambda CDM) = -5.108$. The individual contribution from each dataset in each run for both models is reported in Table \ref{tab:table_2} in Appendix \ref{appx:chi2}. We show the posteriors of $S_8$ and $\Om$ for each dataset in Figure \ref{fig:all s_8_contour_results}, which provides a comparison between MVDM and $\Lambda$CDM models across various datasets. Additionally, we have included the residual plots for both the ACT and ACT+$S_8$ datasets along with the other datasets previously mentioned in Figure \ref{fig:residula_plots}.

We have also performed a comprehensive analysis by integrating the Planck, BAO, and Pantheon+ datasets. The mean and best-fit values of the cosmological parameters derived from this combined analysis are presented in Table \ref{tab:table3}. The results show no significant changes compared to the Planck-only analysis, both in terms of the posterior distributions and the estimated values of the cosmological parameters, as shown in figure \ref{fig:mcmc_results}.

We found the lowest lower limit on the transition redshift $\log_{10} z_t >4.55$ among all combinations of datasets. It is worth noting that the lower bound must be higher than the redshift of recombination to avoid significant impacts on the CMB power spectra. We have also performed an MCMC analysis by considering all of the datasets into account (i.e. both Planck+ACT and Planck+ACT+$S_8$). However, no significant change in the result was observed.

We computed the Akaike Information Criterion (AIC) to investigate any preference of the data for the MVDM model over $\Lambda$CDM. It is a statistical tool for evaluating the relative quality of different models when applied to a specific dataset. The difference in AIC between two models is given by
\begin{equation}\label{eq:AIC}
    \Delta \text{AIC}=  \chi^2_{\rm min, Model} -  \chi^2_{\rm {min},\Lambda \rm CDM} + 2(N_{\rm Model}-N_{\Lambda \rm CDM})\,.
\end{equation} 
The $\Delta \chi^2_{\rm min}$ and $\Delta$AIC for each dataset combination are given in Table \ref{tab:table_aic}. We see that the ACT data exhibits a mild preference for MVDM over $\Lambda$CDM, which is not the case for Planck. This result can also be attributed to the smaller error bars of ACT at high-$\ell$ compared to Planck, as the changes in the power spectrum for MVDM come from the small scales.

\section{Conclusions} \label{conclusion}
In this study, we explored the model of a mass-varying dark matter (MVDM), where it transitions from a relativistic to a nonrelativistic phase in the early Universe, and derived constraints using cosmological observation data. We assumed a phenomenological relation of dark matter mass as a function of redshift without invoking any specific particle physics model. We focused only on fast transitions in this work. The large free streaming velocity in the relativistic phase of DM at early time washes out the structures at small scales and manifests its effect in easing the tension between the amplitude of matter fluctuations ($S_8$) determined from local\,\cite{kidsref, 10.1093/mnras/stt601,10.1093/mnras/stv1154,refId01,refId02,PhysRevD.98.043526} and high redshift probes\,\cite{refId0}, as depicted in Figure \ref{fig:mcmc_results}. Our Bayesian analysis with the MVDM model reveals a slight reduction in $S_8$ when only Planck CMB data is used, and the result does not change significantly when BAO and supernova datasets are included. However, the inclusion of $S_8$ prior, obtained from KIDS1000+BOSS+2dfLenS weak lensing measurements, significantly reduces the value of $S_8$. Furthermore, the model puts a lower bound on the transition redshift and shows a mild preference for the mass of the dark matter to be $23.87 \eV$. Similar $\sim\eV$ order warm dark matter mass has also been predicted in other models \cite{Rogers:2023upm}.

The MVDM model shows new features at high multipoles of the CMB anisotropic power spectra. Because of its relatively larger error bars at high multipoles, Planck-only analysis yields poorer constraints on the DM mass. Since ACT data provides those measurements at higher multipoles with comparatively better precision, our MCMC analysis showed better constraining of mass $M$ compared to Planck. This is also reflected in the $\Delta$AIC calculations. Including prior from KIDS1000+BOSS+2dfLenS weak lensing data even further reduces the $S_8$ as expected. 

Future experiments, such as CMB-S4 \cite{abazajian2016cmbs4,abazajian2022snowmass}, are expected to achieve higher precision measurements at high multipoles, specifically for goals related to de-lensing the inflationary B-modes, constraining $N_{\rm eff}$ and $\Sigma m_\nu$, among others. Such experiments are expected to provide insight into the detailed small-scale physics of the CMB data, which could aid in the detection and proper constraint of the DM mass $M$ of our model. In this regard, a Fisher forecast can also be conducted in the future. Another way to probe such a model is through CMB spectral distortion as shown in \cite{Sarkar_2017}, which might also be detected in future experiments. 

As previously mentioned, we did not invoke any specific particle physics model to explain the mass variation of dark matter. However, considering a particular particle physics model could potentially introduce additional model-dependent effects beyond those reported in this work. The residual plots of CMB power spectra and matter power spectra, as shown in Figure \ref{fig:residula_plots} and referenced in \cite{Das:2023enn}, exhibit oscillatory structures resulting from the additional drag force exerted by the relativistic DM, leading to a phase shift in the baryon acoustic oscillation. A similar phenomenon could occur in the dark sector, too. This model could also be employed to constraint the dark acoustic oscillation\,\cite{10.1093/mnras/stab1116,10.1093/mnrasl/slad107}. Additionally, the present model could be tested by the small-scale Lyman-$\alpha$ forest data in line with recent works \cite{Archidiacono_2019,PhysRevD.100.123520,PhysRevD.88.123515,Murgia_2017,Baur_2017,PhysRevD.98.083540,Palanque-Delabrouille_2020,10.1093/mnras/stab1960}. 

\begin{acknowledgments}
AC thanks Ranjan Laha for the useful discussions regarding this work and some of the future works mentioned above. AD was supported by
Grant Korea NRF-2019R1C1C1010050. SD acknowledges the DST-SERB Government of India grant CRG/2019/006147 for supporting the project
\end{acknowledgments}

\bibliography{main}

\begin{thebibliography}{68}%
\makeatletter
\providecommand \@ifxundefined [1]{%
 \@ifx{#1\undefined}
}%
\providecommand \@ifnum [1]{%
 \ifnum #1\expandafter \@firstoftwo
 \else \expandafter \@secondoftwo
 \fi
}%
\providecommand \@ifx [1]{%
 \ifx #1\expandafter \@firstoftwo
 \else \expandafter \@secondoftwo
 \fi
}%
\providecommand \natexlab [1]{#1}%
\providecommand \enquote  [1]{``#1''}%
\providecommand \bibnamefont  [1]{#1}%
\providecommand \bibfnamefont [1]{#1}%
\providecommand \citenamefont [1]{#1}%
\providecommand \href@noop [0]{\@secondoftwo}%
\providecommand \href [0]{\begingroup \@sanitize@url \@href}%
\providecommand \@href[1]{\@@startlink{#1}\@@href}%
\providecommand \@@href[1]{\endgroup#1\@@endlink}%
\providecommand \@sanitize@url [0]{\catcode `\\12\catcode `\$12\catcode `\&12\catcode `\#12\catcode `\^12\catcode `\_12\catcode `\%12\relax}%
\providecommand \@@startlink[1]{}%
\providecommand \@@endlink[0]{}%
\providecommand \url  [0]{\begingroup\@sanitize@url \@url }%
\providecommand \@url [1]{\endgroup\@href {#1}{\urlprefix }}%
\providecommand \urlprefix  [0]{URL }%
\providecommand \Eprint [0]{\href }%
\providecommand \doibase [0]{https://doi.org/}%
\providecommand \selectlanguage [0]{\@gobble}%
\providecommand \bibinfo  [0]{\@secondoftwo}%
\providecommand \bibfield  [0]{\@secondoftwo}%
\providecommand \translation [1]{[#1]}%
\providecommand \BibitemOpen [0]{}%
\providecommand \bibitemStop [0]{}%
\providecommand \bibitemNoStop [0]{.\EOS\space}%
\providecommand \EOS [0]{\spacefactor3000\relax}%
\providecommand \BibitemShut  [1]{\csname bibitem#1\endcsname}%
\let\auto@bib@innerbib\@empty
\bibitem [{\citenamefont {Aghanim $et~al.$ (Planck~Collaboration)}(2020)}]{refId0}%
  \BibitemOpen
  \bibfield  {author} {\bibinfo {author} {\bibfnamefont {N.}~\bibnamefont {Aghanim $et~al.$ (Planck~Collaboration)}},\ }\href {https://doi.org/10.1051/0004-6361/201833910} {\bibfield  {journal} {\bibinfo  {journal} {A\&A}\ }\textbf {\bibinfo {volume} {641}},\ \bibinfo {pages} {A6} (\bibinfo {year} {2020})}\BibitemShut {NoStop}%
\bibitem [{\citenamefont {Salucci}(2019)}]{Salucci2019}%
  \BibitemOpen
  \bibfield  {author} {\bibinfo {author} {\bibfnamefont {P.}~\bibnamefont {Salucci}},\ }\href {https://doi.org/10.1007/s00159-018-0113-1} {\bibfield  {journal} {\bibinfo  {journal} {The Astronomy and Astrophysics Review}\ }\textbf {\bibinfo {volume} {27}},\ \bibinfo {pages} {2} (\bibinfo {year} {2019})}\BibitemShut {NoStop}%
\bibitem [{\citenamefont {Viel}\ \emph {et~al.}(2013{\natexlab{a}})\citenamefont {Viel}, \citenamefont {Becker}, \citenamefont {Bolton},\ and\ \citenamefont {Haehnelt}}]{PhysRevD.88.043502}%
  \BibitemOpen
  \bibfield  {author} {\bibinfo {author} {\bibfnamefont {M.}~\bibnamefont {Viel}}, \bibinfo {author} {\bibfnamefont {G.~D.}\ \bibnamefont {Becker}}, \bibinfo {author} {\bibfnamefont {J.~S.}\ \bibnamefont {Bolton}},\ and\ \bibinfo {author} {\bibfnamefont {M.~G.}\ \bibnamefont {Haehnelt}},\ }\href {https://doi.org/10.1103/PhysRevD.88.043502} {\bibfield  {journal} {\bibinfo  {journal} {Phys. Rev. D}\ }\textbf {\bibinfo {volume} {88}},\ \bibinfo {pages} {043502} (\bibinfo {year} {2013}{\natexlab{a}})}\BibitemShut {NoStop}%
\bibitem [{\citenamefont {Holm}\ \emph {et~al.}(2022)\citenamefont {Holm}, \citenamefont {Tram},\ and\ \citenamefont {Hannestad}}]{Holm_2022}%
  \BibitemOpen
  \bibfield  {author} {\bibinfo {author} {\bibfnamefont {E.~B.}\ \bibnamefont {Holm}}, \bibinfo {author} {\bibfnamefont {T.}~\bibnamefont {Tram}},\ and\ \bibinfo {author} {\bibfnamefont {S.}~\bibnamefont {Hannestad}},\ }\href {https://doi.org/10.1088/1475-7516/2022/08/044} {\bibfield  {journal} {\bibinfo  {journal} {Journal of Cosmology and Astroparticle Physics}\ }\textbf {\bibinfo {volume} {2022}}\bibinfo  {number} { (08)},\ \bibinfo {pages} {044}}\BibitemShut {NoStop}%
\bibitem [{\citenamefont {Blinov}\ \emph {et~al.}(2020)\citenamefont {Blinov}, \citenamefont {Keith},\ and\ \citenamefont {Hooper}}]{Blinov_2020}%
  \BibitemOpen
\bibfield  {number} {  }\bibfield  {author} {\bibinfo {author} {\bibfnamefont {N.}~\bibnamefont {Blinov}}, \bibinfo {author} {\bibfnamefont {C.}~\bibnamefont {Keith}},\ and\ \bibinfo {author} {\bibfnamefont {D.}~\bibnamefont {Hooper}},\ }\href {https://doi.org/10.1088/1475-7516/2020/06/005} {\bibfield  {journal} {\bibinfo  {journal} {Journal of Cosmology and Astroparticle Physics}\ }\textbf {\bibinfo {volume} {2020}}\bibinfo  {number} { (06)},\ \bibinfo {pages} {005}}\BibitemShut {NoStop}%
\bibitem [{\citenamefont {Alexander}\ \emph {et~al.}(2023)\citenamefont {Alexander}, \citenamefont {Bernardo},\ and\ \citenamefont {Toomey}}]{Alexander_2023}%
  \BibitemOpen
\bibfield  {number} {  }\bibfield  {author} {\bibinfo {author} {\bibfnamefont {S.}~\bibnamefont {Alexander}}, \bibinfo {author} {\bibfnamefont {H.}~\bibnamefont {Bernardo}},\ and\ \bibinfo {author} {\bibfnamefont {M.~W.}\ \bibnamefont {Toomey}},\ }\href {https://doi.org/10.1088/1475-7516/2023/03/037} {\bibfield  {journal} {\bibinfo  {journal} {Journal of Cosmology and Astroparticle Physics}\ }\textbf {\bibinfo {volume} {2023}}\bibinfo  {number} { (03)},\ \bibinfo {pages} {037}}\BibitemShut {NoStop}%
\bibitem [{\citenamefont {Abdalla}\ \emph {et~al.}(2022)\citenamefont {Abdalla} \emph {et~al.}}]{Abdalla:2022yfr}%
  \BibitemOpen
\bibfield  {number} {  }\bibfield  {author} {\bibinfo {author} {\bibfnamefont {E.}~\bibnamefont {Abdalla}} \emph {et~al.},\ }\href {https://doi.org/10.1016/j.jheap.2022.04.002} {\bibfield  {journal} {\bibinfo  {journal} {JHEAp}\ }\textbf {\bibinfo {volume} {34}},\ \bibinfo {pages} {49} (\bibinfo {year} {2022})},\ \Eprint {https://arxiv.org/abs/2203.06142} {arXiv:2203.06142 [astro-ph.CO]} \BibitemShut {NoStop}%
\bibitem [{\citenamefont {{C.Heymans \textit{et al.}}}(2013)}]{10.1093/mnras/stt601}%
  \BibitemOpen
  \bibfield  {author} {\bibinfo {author} {\bibnamefont {{C.Heymans \textit{et al.}}}},\ }\href {https://doi.org/10.1093/mnras/stt601} {\bibfield  {journal} {\bibinfo  {journal} {Monthly Notices of the Royal Astronomical Society}\ }\textbf {\bibinfo {volume} {432}},\ \bibinfo {pages} {2433} (\bibinfo {year} {2013})},\ \Eprint {https://arxiv.org/abs/https://academic.oup.com/mnras/article-pdf/432/3/2433/12627095/stt601.pdf} {https://academic.oup.com/mnras/article-pdf/432/3/2433/12627095/stt601.pdf} \BibitemShut {NoStop}%
\bibitem [{\citenamefont {MacCrann}\ \emph {et~al.}(2015)\citenamefont {MacCrann}, \citenamefont {Zuntz}, \citenamefont {Bridle}, \citenamefont {Jain},\ and\ \citenamefont {Becker}}]{10.1093/mnras/stv1154}%
  \BibitemOpen
  \bibfield  {author} {\bibinfo {author} {\bibfnamefont {N.}~\bibnamefont {MacCrann}}, \bibinfo {author} {\bibfnamefont {J.}~\bibnamefont {Zuntz}}, \bibinfo {author} {\bibfnamefont {S.}~\bibnamefont {Bridle}}, \bibinfo {author} {\bibfnamefont {B.}~\bibnamefont {Jain}},\ and\ \bibinfo {author} {\bibfnamefont {M.~R.}\ \bibnamefont {Becker}},\ }\href {https://doi.org/10.1093/mnras/stv1154} {\bibfield  {journal} {\bibinfo  {journal} {Monthly Notices of the Royal Astronomical Society}\ }\textbf {\bibinfo {volume} {451}},\ \bibinfo {pages} {2877} (\bibinfo {year} {2015})},\ \Eprint {https://arxiv.org/abs/https://academic.oup.com/mnras/article-pdf/451/3/2877/4025570/stv1154.pdf} {https://academic.oup.com/mnras/article-pdf/451/3/2877/4025570/stv1154.pdf} \BibitemShut {NoStop}%
\bibitem [{\citenamefont {Abbott~\textit{et al.}}(2018)}]{PhysRevD.98.043526}%
  \BibitemOpen
  \bibfield  {author} {\bibinfo {author} {\bibfnamefont {T.~M.~C.}\ \bibnamefont {Abbott~\textit{et al.}}} (\bibinfo {collaboration} {Dark Energy Survey Collaboration 1}),\ }\href {https://doi.org/10.1103/PhysRevD.98.043526} {\bibfield  {journal} {\bibinfo  {journal} {Phys. Rev. D}\ }\textbf {\bibinfo {volume} {98}},\ \bibinfo {pages} {043526} (\bibinfo {year} {2018})}\BibitemShut {NoStop}%
\bibitem [{\citenamefont {{H. Hildebrandt \textit{et al.}}}(2020)}]{refId01}%
  \BibitemOpen
  \bibfield  {author} {\bibinfo {author} {\bibnamefont {{H. Hildebrandt \textit{et al.}}}},\ }\href {https://doi.org/10.1051/0004-6361/201834878} {\bibfield  {journal} {\bibinfo  {journal} {A\&A}\ }\textbf {\bibinfo {volume} {633}},\ \bibinfo {pages} {A69} (\bibinfo {year} {2020})}\BibitemShut {NoStop}%
\bibitem [{\citenamefont {{S. Joudaki \textit{et al.} }}(2020)}]{refId02}%
  \BibitemOpen
  \bibfield  {author} {\bibinfo {author} {\bibnamefont {{S. Joudaki \textit{et al.} }}},\ }\href {https://doi.org/10.1051/0004-6361/201936154} {\bibfield  {journal} {\bibinfo  {journal} {A\&A}\ }\textbf {\bibinfo {volume} {638}},\ \bibinfo {pages} {L1} (\bibinfo {year} {2020})}\BibitemShut {NoStop}%
\bibitem [{\citenamefont {Poulin}\ \emph {et~al.}(2018)\citenamefont {Poulin}, \citenamefont {Boddy}, \citenamefont {Bird},\ and\ \citenamefont {Kamionkowski}}]{PhysRevD.97.123504}%
  \BibitemOpen
  \bibfield  {author} {\bibinfo {author} {\bibfnamefont {V.}~\bibnamefont {Poulin}}, \bibinfo {author} {\bibfnamefont {K.~K.}\ \bibnamefont {Boddy}}, \bibinfo {author} {\bibfnamefont {S.}~\bibnamefont {Bird}},\ and\ \bibinfo {author} {\bibfnamefont {M.}~\bibnamefont {Kamionkowski}},\ }\href {https://doi.org/10.1103/PhysRevD.97.123504} {\bibfield  {journal} {\bibinfo  {journal} {Phys. Rev. D}\ }\textbf {\bibinfo {volume} {97}},\ \bibinfo {pages} {123504} (\bibinfo {year} {2018})}\BibitemShut {NoStop}%
\bibitem [{\citenamefont {Enqvist}\ \emph {et~al.}(2015)\citenamefont {Enqvist}, \citenamefont {Nadathur}, \citenamefont {Sekiguchi},\ and\ \citenamefont {Takahashi}}]{Enqvist_2015}%
  \BibitemOpen
  \bibfield  {author} {\bibinfo {author} {\bibfnamefont {K.}~\bibnamefont {Enqvist}}, \bibinfo {author} {\bibfnamefont {S.}~\bibnamefont {Nadathur}}, \bibinfo {author} {\bibfnamefont {T.}~\bibnamefont {Sekiguchi}},\ and\ \bibinfo {author} {\bibfnamefont {T.}~\bibnamefont {Takahashi}},\ }\href {https://doi.org/10.1088/1475-7516/2015/09/067} {\bibfield  {journal} {\bibinfo  {journal} {Journal of Cosmology and Astroparticle Physics}\ }\textbf {\bibinfo {volume} {2015}}\bibinfo  {number} { (09)},\ \bibinfo {pages} {067}}\BibitemShut {NoStop}%
\bibitem [{\citenamefont {Poulin}\ \emph {et~al.}(2016)\citenamefont {Poulin}, \citenamefont {Serpico},\ and\ \citenamefont {Lesgourgues}}]{Poulin_2016}%
  \BibitemOpen
\bibfield  {number} {  }\bibfield  {author} {\bibinfo {author} {\bibfnamefont {V.}~\bibnamefont {Poulin}}, \bibinfo {author} {\bibfnamefont {P.~D.}\ \bibnamefont {Serpico}},\ and\ \bibinfo {author} {\bibfnamefont {J.}~\bibnamefont {Lesgourgues}},\ }\href {https://doi.org/10.1088/1475-7516/2016/08/036} {\bibfield  {journal} {\bibinfo  {journal} {Journal of Cosmology and Astroparticle Physics}\ }\textbf {\bibinfo {volume} {2016}}\bibinfo  {number} { (08)},\ \bibinfo {pages} {036}}\BibitemShut {NoStop}%
\bibitem [{\citenamefont {Vattis}\ \emph {et~al.}(2019)\citenamefont {Vattis}, \citenamefont {Koushiappas},\ and\ \citenamefont {Loeb}}]{PhysRevD.99.121302}%
  \BibitemOpen
\bibfield  {number} {  }\bibfield  {author} {\bibinfo {author} {\bibfnamefont {K.}~\bibnamefont {Vattis}}, \bibinfo {author} {\bibfnamefont {S.~M.}\ \bibnamefont {Koushiappas}},\ and\ \bibinfo {author} {\bibfnamefont {A.}~\bibnamefont {Loeb}},\ }\href {https://doi.org/10.1103/PhysRevD.99.121302} {\bibfield  {journal} {\bibinfo  {journal} {Phys. Rev. D}\ }\textbf {\bibinfo {volume} {99}},\ \bibinfo {pages} {121302} (\bibinfo {year} {2019})}\BibitemShut {NoStop}%
\bibitem [{\citenamefont {Haridasu}\ and\ \citenamefont {Viel}(2020)}]{10.1093/mnras/staa1991}%
  \BibitemOpen
  \bibfield  {author} {\bibinfo {author} {\bibfnamefont {B.~S.}\ \bibnamefont {Haridasu}}\ and\ \bibinfo {author} {\bibfnamefont {M.}~\bibnamefont {Viel}},\ }\href {https://doi.org/10.1093/mnras/staa1991} {\bibfield  {journal} {\bibinfo  {journal} {Monthly Notices of the Royal Astronomical Society}\ }\textbf {\bibinfo {volume} {497}},\ \bibinfo {pages} {1757} (\bibinfo {year} {2020})},\ \Eprint {https://arxiv.org/abs/https://academic.oup.com/mnras/article-pdf/497/2/1757/33562858/staa1991.pdf} {https://academic.oup.com/mnras/article-pdf/497/2/1757/33562858/staa1991.pdf} \BibitemShut {NoStop}%
\bibitem [{\citenamefont {Clark}\ \emph {et~al.}(2021)\citenamefont {Clark}, \citenamefont {Vattis},\ and\ \citenamefont {Koushiappas}}]{PhysRevD.103.043014}%
  \BibitemOpen
  \bibfield  {author} {\bibinfo {author} {\bibfnamefont {S.~J.}\ \bibnamefont {Clark}}, \bibinfo {author} {\bibfnamefont {K.}~\bibnamefont {Vattis}},\ and\ \bibinfo {author} {\bibfnamefont {S.~M.}\ \bibnamefont {Koushiappas}},\ }\href {https://doi.org/10.1103/PhysRevD.103.043014} {\bibfield  {journal} {\bibinfo  {journal} {Phys. Rev. D}\ }\textbf {\bibinfo {volume} {103}},\ \bibinfo {pages} {043014} (\bibinfo {year} {2021})}\BibitemShut {NoStop}%
\bibitem [{\citenamefont {Pandey}\ \emph {et~al.}(2020)\citenamefont {Pandey}, \citenamefont {Karwal},\ and\ \citenamefont {Das}}]{Pandey_2020}%
  \BibitemOpen
  \bibfield  {author} {\bibinfo {author} {\bibfnamefont {K.~L.}\ \bibnamefont {Pandey}}, \bibinfo {author} {\bibfnamefont {T.}~\bibnamefont {Karwal}},\ and\ \bibinfo {author} {\bibfnamefont {S.}~\bibnamefont {Das}},\ }\href {https://doi.org/10.1088/1475-7516/2020/07/026} {\bibfield  {journal} {\bibinfo  {journal} {Journal of Cosmology and Astroparticle Physics}\ }\textbf {\bibinfo {volume} {2020}}\bibinfo  {number} { (07)},\ \bibinfo {pages} {026}}\BibitemShut {NoStop}%
\bibitem [{\citenamefont {Abell\'an}\ \emph {et~al.}(2022)\citenamefont {Abell\'an}, \citenamefont {Murgia}, \citenamefont {Poulin},\ and\ \citenamefont {Lavalle}}]{PhysRevD.105.063525}%
  \BibitemOpen
\bibfield  {number} {  }\bibfield  {author} {\bibinfo {author} {\bibfnamefont {G.~F.}\ \bibnamefont {Abell\'an}}, \bibinfo {author} {\bibfnamefont {R.}~\bibnamefont {Murgia}}, \bibinfo {author} {\bibfnamefont {V.}~\bibnamefont {Poulin}},\ and\ \bibinfo {author} {\bibfnamefont {J.}~\bibnamefont {Lavalle}},\ }\href {https://doi.org/10.1103/PhysRevD.105.063525} {\bibfield  {journal} {\bibinfo  {journal} {Phys. Rev. D}\ }\textbf {\bibinfo {volume} {105}},\ \bibinfo {pages} {063525} (\bibinfo {year} {2022})}\BibitemShut {NoStop}%
\bibitem [{\citenamefont {Abell\'an}\ \emph {et~al.}(2021)\citenamefont {Abell\'an}, \citenamefont {Murgia},\ and\ \citenamefont {Poulin}}]{PhysRevD.104.123533}%
  \BibitemOpen
  \bibfield  {author} {\bibinfo {author} {\bibfnamefont {G.~F.}\ \bibnamefont {Abell\'an}}, \bibinfo {author} {\bibfnamefont {R.}~\bibnamefont {Murgia}},\ and\ \bibinfo {author} {\bibfnamefont {V.}~\bibnamefont {Poulin}},\ }\href {https://doi.org/10.1103/PhysRevD.104.123533} {\bibfield  {journal} {\bibinfo  {journal} {Phys. Rev. D}\ }\textbf {\bibinfo {volume} {104}},\ \bibinfo {pages} {123533} (\bibinfo {year} {2021})}\BibitemShut {NoStop}%
\bibitem [{\citenamefont {Kreisch}\ \emph {et~al.}(2020)\citenamefont {Kreisch}, \citenamefont {Cyr-Racine},\ and\ \citenamefont {Dor\'e}}]{PhysRevD.101.123505}%
  \BibitemOpen
  \bibfield  {author} {\bibinfo {author} {\bibfnamefont {C.~D.}\ \bibnamefont {Kreisch}}, \bibinfo {author} {\bibfnamefont {F.-Y.}\ \bibnamefont {Cyr-Racine}},\ and\ \bibinfo {author} {\bibfnamefont {O.}~\bibnamefont {Dor\'e}},\ }\href {https://doi.org/10.1103/PhysRevD.101.123505} {\bibfield  {journal} {\bibinfo  {journal} {Phys. Rev. D}\ }\textbf {\bibinfo {volume} {101}},\ \bibinfo {pages} {123505} (\bibinfo {year} {2020})}\BibitemShut {NoStop}%
\bibitem [{\citenamefont {Murgia}\ \emph {et~al.}(2016)\citenamefont {Murgia}, \citenamefont {Gariazzo},\ and\ \citenamefont {Fornengo}}]{Murgia_2016}%
  \BibitemOpen
  \bibfield  {author} {\bibinfo {author} {\bibfnamefont {R.}~\bibnamefont {Murgia}}, \bibinfo {author} {\bibfnamefont {S.}~\bibnamefont {Gariazzo}},\ and\ \bibinfo {author} {\bibfnamefont {N.}~\bibnamefont {Fornengo}},\ }\href {https://doi.org/10.1088/1475-7516/2016/04/014} {\bibfield  {journal} {\bibinfo  {journal} {Journal of Cosmology and Astroparticle Physics}\ }\textbf {\bibinfo {volume} {2016}}\bibinfo  {number} { (04)},\ \bibinfo {pages} {014}}\BibitemShut {NoStop}%
\bibitem [{\citenamefont {Archidiacono}\ \emph {et~al.}(2019)\citenamefont {Archidiacono}, \citenamefont {Hooper}, \citenamefont {Murgia}, \citenamefont {Bohr}, \citenamefont {Lesgourgues},\ and\ \citenamefont {Viel}}]{Archidiacono_2019}%
  \BibitemOpen
\bibfield  {number} {  }\bibfield  {author} {\bibinfo {author} {\bibfnamefont {M.}~\bibnamefont {Archidiacono}}, \bibinfo {author} {\bibfnamefont {D.~C.}\ \bibnamefont {Hooper}}, \bibinfo {author} {\bibfnamefont {R.}~\bibnamefont {Murgia}}, \bibinfo {author} {\bibfnamefont {S.}~\bibnamefont {Bohr}}, \bibinfo {author} {\bibfnamefont {J.}~\bibnamefont {Lesgourgues}},\ and\ \bibinfo {author} {\bibfnamefont {M.}~\bibnamefont {Viel}},\ }\href {https://doi.org/10.1088/1475-7516/2019/10/055} {\bibfield  {journal} {\bibinfo  {journal} {Journal of Cosmology and Astroparticle Physics}\ }\textbf {\bibinfo {volume} {2019}}\bibinfo  {number} { (10)},\ \bibinfo {pages} {055}}\BibitemShut {NoStop}%
\bibitem [{\citenamefont {Becker}\ \emph {et~al.}(2021)\citenamefont {Becker}, \citenamefont {Hooper}, \citenamefont {Kahlhoefer}, \citenamefont {Lesgourgues},\ and\ \citenamefont {Schöneberg}}]{Becker_2021}%
  \BibitemOpen
\bibfield  {number} {  }\bibfield  {author} {\bibinfo {author} {\bibfnamefont {N.}~\bibnamefont {Becker}}, \bibinfo {author} {\bibfnamefont {D.~C.}\ \bibnamefont {Hooper}}, \bibinfo {author} {\bibfnamefont {F.}~\bibnamefont {Kahlhoefer}}, \bibinfo {author} {\bibfnamefont {J.}~\bibnamefont {Lesgourgues}},\ and\ \bibinfo {author} {\bibfnamefont {N.}~\bibnamefont {Schöneberg}},\ }\href {https://doi.org/10.1088/1475-7516/2021/02/019} {\bibfield  {journal} {\bibinfo  {journal} {Journal of Cosmology and Astroparticle Physics}\ }\textbf {\bibinfo {volume} {2021}}\bibinfo  {number} { (02)},\ \bibinfo {pages} {019}}\BibitemShut {NoStop}%
\bibitem [{\citenamefont {Kumar}\ and\ \citenamefont {Nunes}(2016)}]{PhysRevD.94.123511}%
  \BibitemOpen
\bibfield  {number} {  }\bibfield  {author} {\bibinfo {author} {\bibfnamefont {S.}~\bibnamefont {Kumar}}\ and\ \bibinfo {author} {\bibfnamefont {R.~C.}\ \bibnamefont {Nunes}},\ }\href {https://doi.org/10.1103/PhysRevD.94.123511} {\bibfield  {journal} {\bibinfo  {journal} {Phys. Rev. D}\ }\textbf {\bibinfo {volume} {94}},\ \bibinfo {pages} {123511} (\bibinfo {year} {2016})}\BibitemShut {NoStop}%
\bibitem [{\citenamefont {Agarwal}\ \emph {et~al.}(2015)\citenamefont {Agarwal}, \citenamefont {Corasaniti}, \citenamefont {Das},\ and\ \citenamefont {Rasera}}]{Agarwal:2014qca}%
  \BibitemOpen
  \bibfield  {author} {\bibinfo {author} {\bibfnamefont {S.}~\bibnamefont {Agarwal}}, \bibinfo {author} {\bibfnamefont {P.~S.}\ \bibnamefont {Corasaniti}}, \bibinfo {author} {\bibfnamefont {S.}~\bibnamefont {Das}},\ and\ \bibinfo {author} {\bibfnamefont {Y.}~\bibnamefont {Rasera}},\ }\href {https://doi.org/10.1103/PhysRevD.92.063502} {\bibfield  {journal} {\bibinfo  {journal} {Phys. Rev. D}\ }\textbf {\bibinfo {volume} {92}},\ \bibinfo {pages} {063502} (\bibinfo {year} {2015})},\ \Eprint {https://arxiv.org/abs/1412.1103} {arXiv:1412.1103 [astro-ph.CO]} \BibitemShut {NoStop}%
\bibitem [{\citenamefont {Das}\ and\ \citenamefont {Weiner}(2011)}]{Das:2006ht}%
  \BibitemOpen
  \bibfield  {author} {\bibinfo {author} {\bibfnamefont {S.}~\bibnamefont {Das}}\ and\ \bibinfo {author} {\bibfnamefont {N.}~\bibnamefont {Weiner}},\ }\href {https://doi.org/10.1103/PhysRevD.84.123511} {\bibfield  {journal} {\bibinfo  {journal} {Phys. Rev. D}\ }\textbf {\bibinfo {volume} {84}},\ \bibinfo {pages} {123511} (\bibinfo {year} {2011})},\ \Eprint {https://arxiv.org/abs/astro-ph/0611353} {arXiv:astro-ph/0611353} \BibitemShut {NoStop}%
\bibitem [{\citenamefont {Chanda}\ and\ \citenamefont {Das}(2017)}]{Chanda:2017coy}%
  \BibitemOpen
  \bibfield  {author} {\bibinfo {author} {\bibfnamefont {P.~K.}\ \bibnamefont {Chanda}}\ and\ \bibinfo {author} {\bibfnamefont {S.}~\bibnamefont {Das}},\ }\href {https://doi.org/10.1103/PhysRevD.95.083008} {\bibfield  {journal} {\bibinfo  {journal} {Phys. Rev. D}\ }\textbf {\bibinfo {volume} {95}},\ \bibinfo {pages} {083008} (\bibinfo {year} {2017})},\ \Eprint {https://arxiv.org/abs/1702.01882} {arXiv:1702.01882 [gr-qc]} \BibitemShut {NoStop}%
\bibitem [{\citenamefont {Ganesan}\ \emph {et~al.}(2024)\citenamefont {Ganesan}, \citenamefont {Chakraborty}, \citenamefont {Ray}, \citenamefont {Das},\ and\ \citenamefont {Banerjee}}]{Ganesan:2024bsf}%
  \BibitemOpen
  \bibfield  {author} {\bibinfo {author} {\bibfnamefont {V.}~\bibnamefont {Ganesan}}, \bibinfo {author} {\bibfnamefont {A.}~\bibnamefont {Chakraborty}}, \bibinfo {author} {\bibfnamefont {T.}~\bibnamefont {Ray}}, \bibinfo {author} {\bibfnamefont {S.}~\bibnamefont {Das}},\ and\ \bibinfo {author} {\bibfnamefont {A.}~\bibnamefont {Banerjee}},\ }\href@noop {} {\  (\bibinfo {year} {2024})},\ \Eprint {https://arxiv.org/abs/2403.14247} {arXiv:2403.14247 [astro-ph.CO]} \BibitemShut {NoStop}%
\bibitem [{\citenamefont {Das}\ \emph {et~al.}(2023)\citenamefont {Das}, \citenamefont {Das},\ and\ \citenamefont {Sethi}}]{Das:2023enn}%
  \BibitemOpen
  \bibfield  {author} {\bibinfo {author} {\bibfnamefont {A.}~\bibnamefont {Das}}, \bibinfo {author} {\bibfnamefont {S.}~\bibnamefont {Das}},\ and\ \bibinfo {author} {\bibfnamefont {S.~K.}\ \bibnamefont {Sethi}},\ }\href {https://doi.org/10.1103/PhysRevD.108.083501} {\bibfield  {journal} {\bibinfo  {journal} {Phys. Rev. D}\ }\textbf {\bibinfo {volume} {108}},\ \bibinfo {pages} {083501} (\bibinfo {year} {2023})},\ \Eprint {https://arxiv.org/abs/2303.17947} {arXiv:2303.17947 [astro-ph.CO]} \BibitemShut {NoStop}%
\bibitem [{\citenamefont {Sarkar}\ \emph {et~al.}(2015)\citenamefont {Sarkar}, \citenamefont {Das},\ and\ \citenamefont {Sethi}}]{Sarkar:2014bca}%
  \BibitemOpen
  \bibfield  {author} {\bibinfo {author} {\bibfnamefont {A.}~\bibnamefont {Sarkar}}, \bibinfo {author} {\bibfnamefont {S.}~\bibnamefont {Das}},\ and\ \bibinfo {author} {\bibfnamefont {S.~K.}\ \bibnamefont {Sethi}},\ }\href {https://doi.org/10.1088/1475-7516/2015/03/004} {\bibfield  {journal} {\bibinfo  {journal} {JCAP}\ }\textbf {\bibinfo {volume} {03}},\ \bibinfo {pages} {004}},\ \Eprint {https://arxiv.org/abs/1410.7129} {arXiv:1410.7129 [astro-ph.CO]} \BibitemShut {NoStop}%
\bibitem [{\citenamefont {Das}\ \emph {et~al.}(2019)\citenamefont {Das}, \citenamefont {Dasgupta},\ and\ \citenamefont {Khatri}}]{Das:2018ons}%
  \BibitemOpen
  \bibfield  {author} {\bibinfo {author} {\bibfnamefont {A.}~\bibnamefont {Das}}, \bibinfo {author} {\bibfnamefont {B.}~\bibnamefont {Dasgupta}},\ and\ \bibinfo {author} {\bibfnamefont {R.}~\bibnamefont {Khatri}},\ }\href {https://doi.org/10.1088/1475-7516/2019/04/018} {\bibfield  {journal} {\bibinfo  {journal} {JCAP}\ }\textbf {\bibinfo {volume} {04}},\ \bibinfo {pages} {018}},\ \Eprint {https://arxiv.org/abs/1811.00028} {arXiv:1811.00028 [astro-ph.CO]} \BibitemShut {NoStop}%
\bibitem [{\citenamefont {Das}\ and\ \citenamefont {Nadler}(2021)}]{PhysRevD.103.043517}%
  \BibitemOpen
  \bibfield  {author} {\bibinfo {author} {\bibfnamefont {S.}~\bibnamefont {Das}}\ and\ \bibinfo {author} {\bibfnamefont {E.~O.}\ \bibnamefont {Nadler}},\ }\href {https://doi.org/10.1103/PhysRevD.103.043517} {\bibfield  {journal} {\bibinfo  {journal} {Phys. Rev. D}\ }\textbf {\bibinfo {volume} {103}},\ \bibinfo {pages} {043517} (\bibinfo {year} {2021})}\BibitemShut {NoStop}%
\bibitem [{\citenamefont {Bhattacharya}\ \emph {et~al.}(2021)\citenamefont {Bhattacharya}, \citenamefont {Das},\ and\ \citenamefont {Dutta}}]{Bhattacharya:2021wnk}%
  \BibitemOpen
  \bibfield  {author} {\bibinfo {author} {\bibfnamefont {S.}~\bibnamefont {Bhattacharya}}, \bibinfo {author} {\bibfnamefont {A.}~\bibnamefont {Das}},\ and\ \bibinfo {author} {\bibfnamefont {K.}~\bibnamefont {Dutta}},\ }\href {https://doi.org/10.1088/1475-7516/2021/10/071} {\bibfield  {journal} {\bibinfo  {journal} {JCAP}\ }\textbf {\bibinfo {volume} {10}},\ \bibinfo {pages} {071}},\ \Eprint {https://arxiv.org/abs/2101.02234} {arXiv:2101.02234 [astro-ph.CO]} \BibitemShut {NoStop}%
\bibitem [{\citenamefont {Chakraborty}\ \emph {et~al.}(2022)\citenamefont {Chakraborty}, \citenamefont {Chanda}, \citenamefont {Pandey},\ and\ \citenamefont {Das}}]{Chakraborty_2022}%
  \BibitemOpen
  \bibfield  {author} {\bibinfo {author} {\bibfnamefont {A.}~\bibnamefont {Chakraborty}}, \bibinfo {author} {\bibfnamefont {P.~K.}\ \bibnamefont {Chanda}}, \bibinfo {author} {\bibfnamefont {K.~L.}\ \bibnamefont {Pandey}},\ and\ \bibinfo {author} {\bibfnamefont {S.}~\bibnamefont {Das}},\ }\href {https://doi.org/10.3847/1538-4357/ac6ddd} {\bibfield  {journal} {\bibinfo  {journal} {The Astrophysical Journal}\ }\textbf {\bibinfo {volume} {932}},\ \bibinfo {pages} {119} (\bibinfo {year} {2022})}\BibitemShut {NoStop}%
\bibitem [{\citenamefont {Heymans~\textit{et al.}}(2021)}]{kidsref}%
  \BibitemOpen
  \bibfield  {author} {\bibinfo {author} {\bibfnamefont {C.}~\bibnamefont {Heymans~\textit{et al.}}},\ }\href {https://doi.org/10.1051/0004-6361/202039063} {\bibfield  {journal} {\bibinfo  {journal} {A\&A}\ }\textbf {\bibinfo {volume} {646}},\ \bibinfo {pages} {A140} (\bibinfo {year} {2021})}\BibitemShut {NoStop}%
\bibitem [{\citenamefont {Peebles}(1993)}]{peebles:1993}%
  \BibitemOpen
  \bibfield  {author} {\bibinfo {author} {\bibfnamefont {P.~J.~E.}\ \bibnamefont {Peebles}},\ }\href@noop {} {\emph {\bibinfo {title} {Principles of Physical Cosmology}}}\ (\bibinfo  {publisher} {Princeton University Press},\ \bibinfo {year} {1993})\BibitemShut {NoStop}%
\bibitem [{\citenamefont {Hu}\ \emph {et~al.}(1998)\citenamefont {Hu}, \citenamefont {Eisenstein},\ and\ \citenamefont {Tegmark}}]{Hu:1997mj}%
  \BibitemOpen
  \bibfield  {author} {\bibinfo {author} {\bibfnamefont {W.}~\bibnamefont {Hu}}, \bibinfo {author} {\bibfnamefont {D.~J.}\ \bibnamefont {Eisenstein}},\ and\ \bibinfo {author} {\bibfnamefont {M.}~\bibnamefont {Tegmark}},\ }\href {https://doi.org/10.1103/PhysRevLett.80.5255} {\bibfield  {journal} {\bibinfo  {journal} {Phys. Rev. Lett.}\ }\textbf {\bibinfo {volume} {80}},\ \bibinfo {pages} {5255} (\bibinfo {year} {1998})},\ \Eprint {https://arxiv.org/abs/astro-ph/9712057} {arXiv:astro-ph/9712057} \BibitemShut {NoStop}%
\bibitem [{\citenamefont {Lesgourgues}\ and\ \citenamefont {Pastor}(2006)}]{Lesgourgues:2006nd}%
  \BibitemOpen
  \bibfield  {author} {\bibinfo {author} {\bibfnamefont {J.}~\bibnamefont {Lesgourgues}}\ and\ \bibinfo {author} {\bibfnamefont {S.}~\bibnamefont {Pastor}},\ }\href {https://doi.org/10.1016/j.physrep.2006.04.001} {\bibfield  {journal} {\bibinfo  {journal} {Phys. Rept.}\ }\textbf {\bibinfo {volume} {429}},\ \bibinfo {pages} {307} (\bibinfo {year} {2006})},\ \Eprint {https://arxiv.org/abs/astro-ph/0603494} {arXiv:astro-ph/0603494} \BibitemShut {NoStop}%
\bibitem [{\citenamefont {Viel}\ \emph {et~al.}(2013{\natexlab{b}})\citenamefont {Viel}, \citenamefont {Becker}, \citenamefont {Bolton},\ and\ \citenamefont {Haehnelt}}]{Viel:2013fqw}%
  \BibitemOpen
  \bibfield  {author} {\bibinfo {author} {\bibfnamefont {M.}~\bibnamefont {Viel}}, \bibinfo {author} {\bibfnamefont {G.~D.}\ \bibnamefont {Becker}}, \bibinfo {author} {\bibfnamefont {J.~S.}\ \bibnamefont {Bolton}},\ and\ \bibinfo {author} {\bibfnamefont {M.~G.}\ \bibnamefont {Haehnelt}},\ }\href {https://doi.org/10.1103/PhysRevD.88.043502} {\bibfield  {journal} {\bibinfo  {journal} {Phys. Rev. D}\ }\textbf {\bibinfo {volume} {88}},\ \bibinfo {pages} {043502} (\bibinfo {year} {2013}{\natexlab{b}})},\ \Eprint {https://arxiv.org/abs/1306.2314} {arXiv:1306.2314 [astro-ph.CO]} \BibitemShut {NoStop}%
\bibitem [{\citenamefont {Abazajian}\ \emph {et~al.}(2015)\citenamefont {Abazajian} \emph {et~al.}}]{TopicalConvenersKNAbazajianJECarlstromATLee:2013bxd}%
  \BibitemOpen
  \bibfield  {author} {\bibinfo {author} {\bibfnamefont {K.~N.}\ \bibnamefont {Abazajian}} \emph {et~al.} (\bibinfo {collaboration} {Topical Conveners: K.N. Abazajian, J.E. Carlstrom, A.T. Lee}),\ }\href {https://doi.org/10.1016/j.astropartphys.2014.05.014} {\bibfield  {journal} {\bibinfo  {journal} {Astropart. Phys.}\ }\textbf {\bibinfo {volume} {63}},\ \bibinfo {pages} {66} (\bibinfo {year} {2015})},\ \Eprint {https://arxiv.org/abs/1309.5383} {arXiv:1309.5383 [astro-ph.CO]} \BibitemShut {NoStop}%
\bibitem [{\citenamefont {Aiola}\ \emph {et~al.}(2020)\citenamefont {Aiola} \emph {et~al.}}]{Aiola_2020}%
  \BibitemOpen
  \bibfield  {author} {\bibinfo {author} {\bibfnamefont {S.}~\bibnamefont {Aiola}} \emph {et~al.},\ }\href {https://doi.org/10.1088/1475-7516/2020/12/047} {\bibfield  {journal} {\bibinfo  {journal} {Journal of Cosmology and Astroparticle Physics}\ }\textbf {\bibinfo {volume} {2020}}\bibinfo  {number} { (12)},\ \bibinfo {pages} {047}}\BibitemShut {NoStop}%
\bibitem [{\citenamefont {Louis}\ \emph {et~al.}(2017)\citenamefont {Louis} \emph {et~al.}}]{Louis_2017}%
  \BibitemOpen
\bibfield  {number} {  }\bibfield  {author} {\bibinfo {author} {\bibfnamefont {T.}~\bibnamefont {Louis}} \emph {et~al.},\ }\href {https://doi.org/10.1088/1475-7516/2017/06/031} {\bibfield  {journal} {\bibinfo  {journal} {Journal of Cosmology and Astroparticle Physics}\ }\textbf {\bibinfo {volume} {2017}}\bibinfo  {number} { (06)},\ \bibinfo {pages} {031}}\BibitemShut {NoStop}%
\bibitem [{\citenamefont {Amon}\ \emph {et~al.}(2022)\citenamefont {Amon} \emph {et~al.}}]{PhysRevD.105.023514}%
  \BibitemOpen
\bibfield  {number} {  }\bibfield  {author} {\bibinfo {author} {\bibfnamefont {A.}~\bibnamefont {Amon}} \emph {et~al.} (\bibinfo {collaboration} {DES Collaboration}),\ }\href {https://doi.org/10.1103/PhysRevD.105.023514} {\bibfield  {journal} {\bibinfo  {journal} {Phys. Rev. D}\ }\textbf {\bibinfo {volume} {105}},\ \bibinfo {pages} {023514} (\bibinfo {year} {2022})}\BibitemShut {NoStop}%
\bibitem [{\citenamefont {{Asgari, Marika}}\ \emph {et~al.}(2021)\citenamefont {{Asgari, Marika}} \emph {et~al.}}]{refId0kid}%
  \BibitemOpen
  \bibfield  {author} {\bibinfo {author} {\bibnamefont {{Asgari, Marika}}} \emph {et~al.},\ }\href {https://doi.org/10.1051/0004-6361/202039070} {\bibfield  {journal} {\bibinfo  {journal} {A\&A}\ }\textbf {\bibinfo {volume} {645}},\ \bibinfo {pages} {A104} (\bibinfo {year} {2021})}\BibitemShut {NoStop}%
\bibitem [{\citenamefont {Lesgourgues}(2011)}]{lesgourgues2011cosmic}%
  \BibitemOpen
  \bibfield  {author} {\bibinfo {author} {\bibfnamefont {J.}~\bibnamefont {Lesgourgues}},\ }\href@noop {} {\bibinfo {title} {The cosmic linear anisotropy solving system (class) i: Overview}} (\bibinfo {year} {2011}),\ \Eprint {https://arxiv.org/abs/1104.2932} {arXiv:1104.2932 [astro-ph.IM]} \BibitemShut {NoStop}%
\bibitem [{\citenamefont {Beutler}\ \emph {et~al.}(2011)\citenamefont {Beutler} \emph {et~al.}}]{10.1111/j.1365-2966.2011.19250.x}%
  \BibitemOpen
  \bibfield  {author} {\bibinfo {author} {\bibfnamefont {F.}~\bibnamefont {Beutler}} \emph {et~al.},\ }\href {https://doi.org/10.1111/j.1365-2966.2011.19250.x} {\bibfield  {journal} {\bibinfo  {journal} {Monthly Notices of the Royal Astronomical Society}\ }\textbf {\bibinfo {volume} {416}},\ \bibinfo {pages} {3017} (\bibinfo {year} {2011})},\ \Eprint {https://arxiv.org/abs/https://academic.oup.com/mnras/article-pdf/416/4/3017/2985042/mnras0416-3017.pdf} {https://academic.oup.com/mnras/article-pdf/416/4/3017/2985042/mnras0416-3017.pdf} \BibitemShut {NoStop}%
\bibitem [{\citenamefont {Ross}\ \emph {et~al.}(2015)\citenamefont {Ross} \emph {et~al.}}]{10.1093/mnras/stv154}%
  \BibitemOpen
  \bibfield  {author} {\bibinfo {author} {\bibfnamefont {A.~J.}\ \bibnamefont {Ross}} \emph {et~al.},\ }\href {https://doi.org/10.1093/mnras/stv154} {\bibfield  {journal} {\bibinfo  {journal} {Monthly Notices of the Royal Astronomical Society}\ }\textbf {\bibinfo {volume} {449}},\ \bibinfo {pages} {835} (\bibinfo {year} {2015})},\ \Eprint {https://arxiv.org/abs/https://academic.oup.com/mnras/article-pdf/449/1/835/13767551/stv154.pdf} {https://academic.oup.com/mnras/article-pdf/449/1/835/13767551/stv154.pdf} \BibitemShut {NoStop}%
\bibitem [{\citenamefont {Alam}\ \emph {et~al.}(2017)\citenamefont {Alam} \emph {et~al.}}]{10.1093/mnras/stx721}%
  \BibitemOpen
  \bibfield  {author} {\bibinfo {author} {\bibfnamefont {S.}~\bibnamefont {Alam}} \emph {et~al.},\ }\href {https://doi.org/10.1093/mnras/stx721} {\bibfield  {journal} {\bibinfo  {journal} {Monthly Notices of the Royal Astronomical Society}\ }\textbf {\bibinfo {volume} {470}},\ \bibinfo {pages} {2617} (\bibinfo {year} {2017})},\ \Eprint {https://arxiv.org/abs/https://academic.oup.com/mnras/article-pdf/470/3/2617/18315003/stx721.pdf} {https://academic.oup.com/mnras/article-pdf/470/3/2617/18315003/stx721.pdf} \BibitemShut {NoStop}%
\bibitem [{\citenamefont {Brout}\ \emph {et~al.}(2022)\citenamefont {Brout} \emph {et~al.}}]{Brout_2022}%
  \BibitemOpen
  \bibfield  {author} {\bibinfo {author} {\bibfnamefont {D.}~\bibnamefont {Brout}} \emph {et~al.},\ }\href {https://doi.org/10.3847/1538-4357/ac8e04} {\bibfield  {journal} {\bibinfo  {journal} {The Astrophysical Journal}\ }\textbf {\bibinfo {volume} {938}},\ \bibinfo {pages} {110} (\bibinfo {year} {2022})}\BibitemShut {NoStop}%
\bibitem [{\citenamefont {Brinckmann}\ and\ \citenamefont {Lesgourgues}(2018)}]{brinckmann2018montepython}%
  \BibitemOpen
  \bibfield  {author} {\bibinfo {author} {\bibfnamefont {T.}~\bibnamefont {Brinckmann}}\ and\ \bibinfo {author} {\bibfnamefont {J.}~\bibnamefont {Lesgourgues}},\ }\href@noop {} {\bibinfo {title} {Montepython 3: boosted mcmc sampler and other features}} (\bibinfo {year} {2018}),\ \Eprint {https://arxiv.org/abs/1804.07261} {arXiv:1804.07261 [astro-ph.CO]} \BibitemShut {NoStop}%
\bibitem [{\citenamefont {James}\ and\ \citenamefont {Roos}(1975)}]{JAMES1975343}%
  \BibitemOpen
  \bibfield  {author} {\bibinfo {author} {\bibfnamefont {F.}~\bibnamefont {James}}\ and\ \bibinfo {author} {\bibfnamefont {M.}~\bibnamefont {Roos}},\ }\href {https://doi.org/https://doi.org/10.1016/0010-4655(75)90039-9} {\bibfield  {journal} {\bibinfo  {journal} {Computer Physics Communications}\ }\textbf {\bibinfo {volume} {10}},\ \bibinfo {pages} {343} (\bibinfo {year} {1975})}\BibitemShut {NoStop}%
\bibitem [{\citenamefont {Lewis}\ \emph {et~al.}(2000)\citenamefont {Lewis}, \citenamefont {Challinor},\ and\ \citenamefont {Lasenby}}]{Lewis_2000}%
  \BibitemOpen
  \bibfield  {author} {\bibinfo {author} {\bibfnamefont {A.}~\bibnamefont {Lewis}}, \bibinfo {author} {\bibfnamefont {A.}~\bibnamefont {Challinor}},\ and\ \bibinfo {author} {\bibfnamefont {A.}~\bibnamefont {Lasenby}},\ }\href {https://doi.org/10.1086/309179} {\bibfield  {journal} {\bibinfo  {journal} {The Astrophysical Journal}\ }\textbf {\bibinfo {volume} {538}},\ \bibinfo {pages} {473} (\bibinfo {year} {2000})}\BibitemShut {NoStop}%
\bibitem [{\citenamefont {Gelman}\ and\ \citenamefont {Rubin}(1992)}]{10.1214/ss/1177011136}%
  \BibitemOpen
  \bibfield  {author} {\bibinfo {author} {\bibfnamefont {A.}~\bibnamefont {Gelman}}\ and\ \bibinfo {author} {\bibfnamefont {D.~B.}\ \bibnamefont {Rubin}},\ }\href {https://doi.org/10.1214/ss/1177011136} {\bibfield  {journal} {\bibinfo  {journal} {Statistical Science}\ }\textbf {\bibinfo {volume} {7}},\ \bibinfo {pages} {457 } (\bibinfo {year} {1992})}\BibitemShut {NoStop}%
\bibitem [{\citenamefont {Rogers}\ and\ \citenamefont {Poulin}(2023)}]{Rogers:2023upm}%
  \BibitemOpen
  \bibfield  {author} {\bibinfo {author} {\bibfnamefont {K.~K.}\ \bibnamefont {Rogers}}\ and\ \bibinfo {author} {\bibfnamefont {V.}~\bibnamefont {Poulin}},\ }\href@noop {} {\  (\bibinfo {year} {2023})},\ \Eprint {https://arxiv.org/abs/2311.16377} {arXiv:2311.16377 [astro-ph.CO]} \BibitemShut {NoStop}%
\bibitem [{\citenamefont {Abazajian}\ \emph {et~al.}(2016)\citenamefont {Abazajian} \emph {et~al.}}]{abazajian2016cmbs4}%
  \BibitemOpen
  \bibfield  {author} {\bibinfo {author} {\bibfnamefont {K.~N.}\ \bibnamefont {Abazajian}} \emph {et~al.},\ }\href@noop {} {\bibinfo {title} {Cmb-s4 science book, first edition}} (\bibinfo {year} {2016}),\ \Eprint {https://arxiv.org/abs/1610.02743} {arXiv:1610.02743 [astro-ph.CO]} \BibitemShut {NoStop}%
\bibitem [{\citenamefont {Abazajian}\ \emph {et~al.}(2022)\citenamefont {Abazajian} \emph {et~al.}}]{abazajian2022snowmass}%
  \BibitemOpen
  \bibfield  {author} {\bibinfo {author} {\bibfnamefont {K.}~\bibnamefont {Abazajian}} \emph {et~al.},\ }\href@noop {} {\bibinfo {title} {Snowmass 2021 cmb-s4 white paper}} (\bibinfo {year} {2022}),\ \Eprint {https://arxiv.org/abs/2203.08024} {arXiv:2203.08024 [astro-ph.CO]} \BibitemShut {NoStop}%
\bibitem [{\citenamefont {Sarkar}\ \emph {et~al.}(2017)\citenamefont {Sarkar}, \citenamefont {Sethi},\ and\ \citenamefont {Das}}]{Sarkar_2017}%
  \BibitemOpen
  \bibfield  {author} {\bibinfo {author} {\bibfnamefont {A.}~\bibnamefont {Sarkar}}, \bibinfo {author} {\bibfnamefont {S.~K.}\ \bibnamefont {Sethi}},\ and\ \bibinfo {author} {\bibfnamefont {S.}~\bibnamefont {Das}},\ }\href {https://doi.org/10.1088/1475-7516/2017/07/012} {\bibfield  {journal} {\bibinfo  {journal} {Journal of Cosmology and Astroparticle Physics}\ }\textbf {\bibinfo {volume} {2017}}\bibinfo  {number} { (07)},\ \bibinfo {pages} {012}}\BibitemShut {NoStop}%
\bibitem [{\citenamefont {Schaeffer}\ and\ \citenamefont {Schneider}(2021)}]{10.1093/mnras/stab1116}%
  \BibitemOpen
\bibfield  {number} {  }\bibfield  {author} {\bibinfo {author} {\bibfnamefont {T.}~\bibnamefont {Schaeffer}}\ and\ \bibinfo {author} {\bibfnamefont {A.}~\bibnamefont {Schneider}},\ }\href {https://doi.org/10.1093/mnras/stab1116} {\bibfield  {journal} {\bibinfo  {journal} {Monthly Notices of the Royal Astronomical Society}\ }\textbf {\bibinfo {volume} {504}},\ \bibinfo {pages} {3773} (\bibinfo {year} {2021})},\ \Eprint {https://arxiv.org/abs/https://academic.oup.com/mnras/article-pdf/504/3/3773/37896468/stab1116.pdf} {https://academic.oup.com/mnras/article-pdf/504/3/3773/37896468/stab1116.pdf} \BibitemShut {NoStop}%
\bibitem [{\citenamefont {Parashari}\ and\ \citenamefont {Laha}(2023)}]{10.1093/mnrasl/slad107}%
  \BibitemOpen
  \bibfield  {author} {\bibinfo {author} {\bibfnamefont {P.}~\bibnamefont {Parashari}}\ and\ \bibinfo {author} {\bibfnamefont {R.}~\bibnamefont {Laha}},\ }\href {https://doi.org/10.1093/mnrasl/slad107} {\bibfield  {journal} {\bibinfo  {journal} {Monthly Notices of the Royal Astronomical Society: Letters}\ }\textbf {\bibinfo {volume} {526}},\ \bibinfo {pages} {L63} (\bibinfo {year} {2023})},\ \Eprint {https://arxiv.org/abs/https://academic.oup.com/mnrasl/article-pdf/526/1/L63/54609777/slad107.pdf} {https://academic.oup.com/mnrasl/article-pdf/526/1/L63/54609777/slad107.pdf} \BibitemShut {NoStop}%
\bibitem [{\citenamefont {Miller}\ \emph {et~al.}(2019)\citenamefont {Miller}, \citenamefont {Erickcek},\ and\ \citenamefont {Murgia}}]{PhysRevD.100.123520}%
  \BibitemOpen
  \bibfield  {author} {\bibinfo {author} {\bibfnamefont {C.}~\bibnamefont {Miller}}, \bibinfo {author} {\bibfnamefont {A.~L.}\ \bibnamefont {Erickcek}},\ and\ \bibinfo {author} {\bibfnamefont {R.}~\bibnamefont {Murgia}},\ }\href {https://doi.org/10.1103/PhysRevD.100.123520} {\bibfield  {journal} {\bibinfo  {journal} {Phys. Rev. D}\ }\textbf {\bibinfo {volume} {100}},\ \bibinfo {pages} {123520} (\bibinfo {year} {2019})}\BibitemShut {NoStop}%
\bibitem [{\citenamefont {Wang}\ \emph {et~al.}(2013)\citenamefont {Wang}, \citenamefont {Croft}, \citenamefont {Peter}, \citenamefont {Zentner},\ and\ \citenamefont {Purcell}}]{PhysRevD.88.123515}%
  \BibitemOpen
  \bibfield  {author} {\bibinfo {author} {\bibfnamefont {M.-Y.}\ \bibnamefont {Wang}}, \bibinfo {author} {\bibfnamefont {R.~A.~C.}\ \bibnamefont {Croft}}, \bibinfo {author} {\bibfnamefont {A.~H.~G.}\ \bibnamefont {Peter}}, \bibinfo {author} {\bibfnamefont {A.~R.}\ \bibnamefont {Zentner}},\ and\ \bibinfo {author} {\bibfnamefont {C.~W.}\ \bibnamefont {Purcell}},\ }\href {https://doi.org/10.1103/PhysRevD.88.123515} {\bibfield  {journal} {\bibinfo  {journal} {Phys. Rev. D}\ }\textbf {\bibinfo {volume} {88}},\ \bibinfo {pages} {123515} (\bibinfo {year} {2013})}\BibitemShut {NoStop}%
\bibitem [{\citenamefont {Murgia}\ \emph {et~al.}(2017)\citenamefont {Murgia}, \citenamefont {Merle}, \citenamefont {Viel}, \citenamefont {Totzauer},\ and\ \citenamefont {Schneider}}]{Murgia_2017}%
  \BibitemOpen
  \bibfield  {author} {\bibinfo {author} {\bibfnamefont {R.}~\bibnamefont {Murgia}}, \bibinfo {author} {\bibfnamefont {A.}~\bibnamefont {Merle}}, \bibinfo {author} {\bibfnamefont {M.}~\bibnamefont {Viel}}, \bibinfo {author} {\bibfnamefont {M.}~\bibnamefont {Totzauer}},\ and\ \bibinfo {author} {\bibfnamefont {A.}~\bibnamefont {Schneider}},\ }\href {https://doi.org/10.1088/1475-7516/2017/11/046} {\bibfield  {journal} {\bibinfo  {journal} {Journal of Cosmology and Astroparticle Physics}\ }\textbf {\bibinfo {volume} {2017}}\bibinfo  {number} { (11)},\ \bibinfo {pages} {046}}\BibitemShut {NoStop}%
\bibitem [{\citenamefont {Baur}\ \emph {et~al.}(2017)\citenamefont {Baur}, \citenamefont {Palanque-Delabrouille}, \citenamefont {Yèche}, \citenamefont {Boyarsky}, \citenamefont {Ruchayskiy}, \citenamefont {Éric Armengaud},\ and\ \citenamefont {Lesgourgues}}]{Baur_2017}%
  \BibitemOpen
\bibfield  {number} {  }\bibfield  {author} {\bibinfo {author} {\bibfnamefont {J.}~\bibnamefont {Baur}}, \bibinfo {author} {\bibfnamefont {N.}~\bibnamefont {Palanque-Delabrouille}}, \bibinfo {author} {\bibfnamefont {C.}~\bibnamefont {Yèche}}, \bibinfo {author} {\bibfnamefont {A.}~\bibnamefont {Boyarsky}}, \bibinfo {author} {\bibfnamefont {O.}~\bibnamefont {Ruchayskiy}}, \bibinfo {author} {\bibnamefont {Éric Armengaud}},\ and\ \bibinfo {author} {\bibfnamefont {J.}~\bibnamefont {Lesgourgues}},\ }\href {https://doi.org/10.1088/1475-7516/2017/12/013} {\bibfield  {journal} {\bibinfo  {journal} {Journal of Cosmology and Astroparticle Physics}\ }\textbf {\bibinfo {volume} {2017}}\bibinfo  {number} { (12)},\ \bibinfo {pages} {013}}\BibitemShut {NoStop}%
\bibitem [{\citenamefont {Murgia}\ \emph {et~al.}(2018)\citenamefont {Murgia}, \citenamefont {Ir\ifmmode \check{s}\else \v{s}\fi{}i\ifmmode~\check{c}\else \v{c}\fi{}},\ and\ \citenamefont {Viel}}]{PhysRevD.98.083540}%
  \BibitemOpen
\bibfield  {number} {  }\bibfield  {author} {\bibinfo {author} {\bibfnamefont {R.}~\bibnamefont {Murgia}}, \bibinfo {author} {\bibfnamefont {V.}~\bibnamefont {Ir\ifmmode \check{s}\else \v{s}\fi{}i\ifmmode~\check{c}\else \v{c}\fi{}}},\ and\ \bibinfo {author} {\bibfnamefont {M.}~\bibnamefont {Viel}},\ }\href {https://doi.org/10.1103/PhysRevD.98.083540} {\bibfield  {journal} {\bibinfo  {journal} {Phys. Rev. D}\ }\textbf {\bibinfo {volume} {98}},\ \bibinfo {pages} {083540} (\bibinfo {year} {2018})}\BibitemShut {NoStop}%
\bibitem [{\citenamefont {Palanque-Delabrouille}\ \emph {et~al.}(2020)\citenamefont {Palanque-Delabrouille}, \citenamefont {Yèche}, \citenamefont {Schöneberg}, \citenamefont {Lesgourgues}, \citenamefont {Walther}, \citenamefont {Chabanier},\ and\ \citenamefont {Armengaud}}]{Palanque-Delabrouille_2020}%
  \BibitemOpen
  \bibfield  {author} {\bibinfo {author} {\bibfnamefont {N.}~\bibnamefont {Palanque-Delabrouille}}, \bibinfo {author} {\bibfnamefont {C.}~\bibnamefont {Yèche}}, \bibinfo {author} {\bibfnamefont {N.}~\bibnamefont {Schöneberg}}, \bibinfo {author} {\bibfnamefont {J.}~\bibnamefont {Lesgourgues}}, \bibinfo {author} {\bibfnamefont {M.}~\bibnamefont {Walther}}, \bibinfo {author} {\bibfnamefont {S.}~\bibnamefont {Chabanier}},\ and\ \bibinfo {author} {\bibfnamefont {E.}~\bibnamefont {Armengaud}},\ }\href {https://doi.org/10.1088/1475-7516/2020/04/038} {\bibfield  {journal} {\bibinfo  {journal} {Journal of Cosmology and Astroparticle Physics}\ }\textbf {\bibinfo {volume} {2020}}\bibinfo  {number} { (04)},\ \bibinfo {pages} {038}}\BibitemShut {NoStop}%
\bibitem [{\citenamefont {Enzi}\ \emph {et~al.}(2021)\citenamefont {Enzi}, \citenamefont {Murgia}, \citenamefont {Newton}, \citenamefont {Vegetti}, \citenamefont {Frenk}, \citenamefont {Viel}, \citenamefont {Cautun}, \citenamefont {Fassnacht}, \citenamefont {Auger}, \citenamefont {Despali}, \citenamefont {McKean}, \citenamefont {Koopmans},\ and\ \citenamefont {Lovell}}]{10.1093/mnras/stab1960}%
  \BibitemOpen
\bibfield  {number} {  }\bibfield  {author} {\bibinfo {author} {\bibfnamefont {W.}~\bibnamefont {Enzi}}, \bibinfo {author} {\bibfnamefont {R.}~\bibnamefont {Murgia}}, \bibinfo {author} {\bibfnamefont {O.}~\bibnamefont {Newton}}, \bibinfo {author} {\bibfnamefont {S.}~\bibnamefont {Vegetti}}, \bibinfo {author} {\bibfnamefont {C.}~\bibnamefont {Frenk}}, \bibinfo {author} {\bibfnamefont {M.}~\bibnamefont {Viel}}, \bibinfo {author} {\bibfnamefont {M.}~\bibnamefont {Cautun}}, \bibinfo {author} {\bibfnamefont {C.~D.}\ \bibnamefont {Fassnacht}}, \bibinfo {author} {\bibfnamefont {M.}~\bibnamefont {Auger}}, \bibinfo {author} {\bibfnamefont {G.}~\bibnamefont {Despali}}, \bibinfo {author} {\bibfnamefont {J.}~\bibnamefont {McKean}}, \bibinfo {author} {\bibfnamefont {L.~V.~E.}\ \bibnamefont {Koopmans}},\ and\ \bibinfo {author} {\bibfnamefont {M.}~\bibnamefont {Lovell}},\ }\href {https://doi.org/10.1093/mnras/stab1960} {\bibfield  {journal} {\bibinfo  {journal} {Monthly Notices of the Royal Astronomical Society}\ }\textbf
  {\bibinfo {volume} {506}},\ \bibinfo {pages} {5848} (\bibinfo {year} {2021})},\ \Eprint {https://arxiv.org/abs/https://academic.oup.com/mnras/article-pdf/506/4/5848/39729079/stab1960.pdf} {https://academic.oup.com/mnras/article-pdf/506/4/5848/39729079/stab1960.pdf} \BibitemShut {NoStop}%
\end{thebibliography}%
\bibliographystyle{apsrev4-2}

\appendix
\section{$\chi^2_{min}$ Per Experiment} \label{appx:chi2}
\begin{table*}
\caption{\label{tab:table_1} Best-fit $\chi^2_{\rm min}$ per experiment (and total) for both the model}
\begin{ruledtabular}
\begin{tabular}{ccccccc}
 Experiment&\multicolumn{3}{c}{$\Lambda$CDM}&\multicolumn{3}{c}{MVDM}\\[1ex] \hline\rule{0pt}{1.2\normalbaselineskip}

Planck high-$\ell$ TT,TE,EE&2353.36&2354.37&2358.06&2350.87&2352.13&2353.6 \\[1ex]
Planck low-$\ell$ EE&395.75&395.85&395.8&396.15&396.82&395.84 \\[1ex]
Planck low-$\ell$ TT&22.67&22.70&21.9&23&23.35&23.6 \\[1ex]\hline \rule{0pt}{1.2\normalbaselineskip}
KiDS/BOSS/2dFGS&--&--&2.38&--&--&1.29\\[1ex] \hline \rule{0pt}{1.2\normalbaselineskip}
BAO-boss-dr12&--&4.10&--&--&4.09&--\\[1ex]
BAO-smallz-2014&--&1.31&--&--&1.32&--\\[1ex]
Pantheon+&--&1411.59&--&--&1411.95&--\\[1ex] \hline \rule{0pt}{1.2\normalbaselineskip}
Total&2771.78&4189.92&2778.14&2770.55&4189.66&2774.33 \\[1ex]
\end{tabular}
\end{ruledtabular}
\end{table*}

\begin{table*}
\caption{\label{tab:table_2} Best-fit $\chi^2_{\rm min}$ per experiment (and total) for both the model}
\begin{ruledtabular}
\begin{tabular}{ccccc}
 Experiment&\multicolumn{2}{c}{$\Lambda$CDM}&\multicolumn{2}{c}{MVDM}\\[1ex] \hline\rule{0pt}{1.2\normalbaselineskip}

ACTPol lite DR4&280.0462&281.1964&275.7791&276.5869 \\[1ex]
$\tau_{\rm reio}$ prior&0.0033&395.8&0.0041&0.1261 \\[1ex]\hline \rule{0pt}{1.2\normalbaselineskip}
KiDS/BOSS/2dFGS&--&0.6766&--&0.0729\\[1ex] \hline \rule{0pt}{1.2\normalbaselineskip}
Total&280.0495&281.8939&275.7832&276.7859 \\[1ex]
\end{tabular}
\end{ruledtabular}
\end{table*}

\end{document}